\documentclass[structabstract]{aa}  
\usepackage{amssymb}
\usepackage[usenames, dvipsnames]{color}
\usepackage{graphicx}
\usepackage{natbib}
\usepackage[rightcaption]{sidecap}
\usepackage{txfonts}
\usepackage{units}
\usepackage{url}
\newcommand\tsp{\mbox{$\;\!$}}
\graphicspath{{//home/meetu/article3p/manuscript/astroph/}}
\usepackage{hyperref}
\hypersetup{
    final=true,
    pageanchor=true,
    colorlinks=true,
    breaklinks=true,
    linkcolor=blue,
    citecolor=blue,
    urlcolor=blue,
    pdfpagemode=UseNone,
    pdftitle={Horizontal flow fields observed in Hinode G-band images.
        IV. Statisctical study of dynamical environment around pores.},
    pdfauthor={M. Verma and C. Denker},
    pdfsubject={Solar Physics},
    pdfkeywords={Sun: chromosphere, Sun: photosphere, Sun: surface magnetism,
    Sun: sunspots, Techniques: image processing, Methods: data analysis}}
\newcommand\phn{\phantom{0}}
\newcommand\phnn{\phantom{00}}
\newcommand\phs{\phantom{$_0$}}

\begin{document}

%###############################################################################
%#
%#    TITLE
%#
%###############################################################################

\title{Horizontal flow fields observed in Hinode G-band images}

\subtitle{IV.\ Statistical properties of the dynamical environment around pores}

\titlerunning{Horizontal flow fields observed in Hinode G-band images. IV.}

\author{%
    M.\ Verma\inst{1,2} \and
    C.\ Denker\inst{1}}

\institute{%
    Leibniz-Institut f\"ur Astrophysik Potsdam (AIP),
    An der Sternwarte 16,
    14482 Potsdam,
    Germany\\
    \email{\textcolor{blue}{cdenker@aip.de}}
\and
    Max-Planck-Institut f{\"u}r Sonnensystemforschung,
    Max-Planck-Stra{\ss}e 2,
    37191 Katlenburg-Lindau,
    Germany\\
    \email{\textcolor{blue}{verma@mps.mpg.de}}}

\date{Received August 12, 2013; accepted January 16, 2014}

\abstract 
% context heading (optional)
{Solar pores are penumbra-lacking magnetic
features, which mark two important transitions in the spectrum of
magnetohydrodynamic processes: (1) the magnetic field becomes sufficiently
strong to suppress the convective energy transport and (2) at some critical
point some pores develop a penumbra and become sunspots.}
% aims heading (mandatory)
{The purpose of this statistical study is to comprehensively
describe solar pores in terms of their size, perimeter, shape, photometric
properties, and horizontal proper motions. The seeing-free and uniform data of
the Japanese Hinode mission provide an opportunity to compare flow fields in the
vicinity of pores in different environments and at various stages of their
evolution.}
% methods heading (mandatory)
{The extensive database of high-resolution G-band images observed with the
Hinode Solar Optical Telescope (SOT) is a unique resource to derive statistical
properties of pores using advanced digital image processing techniques. The
study is based on two data sets: (1) Photometric and morphological properties
inferred from single G-band images cover almost seven years from 2006 October~25
to 2013 August~31. (2) Horizontal flow fields have been
derived from 356 one-hour sequences of G-band images using local correlation
tracking (LCT) for a shorter period of time from 2006 November~3 to 2008
January~6 comprising 13 active regions.}
% results heading (mandatory)
{A total of 7643/2863 (single/time-averaged) pores builds the foundation of the
statistical analysis. Pores are preferentially observed at low latitudes in the
southern hemisphere during the deep minimum of solar cycle No.~23. This
imbalance reverses during the rise of cycle No.~24, when the pores migrate from
high to low latitudes. Pores are rarely encountered in quiet-Sun G-band images,
and only about 10\% of pores exists in isolation. In general, pores do not
exhibit a circular shape. Typical aspect ratios of the semi-major and -minor
axes are 3:2 when ellipses are fitted to pores. Smaller pores (more than
two-thirds are smaller than 5~Mm$^2$) tend to be more circular, and their
boundaries are less corrugated. Both area and perimeter length of pores obey
log-normal frequency distributions. The frequency distribution of the intensity
can be reproduced by two Gaussians representing dark and bright components.
Bright features resembling umbral dots and even light-bridges cover about 20\%
of the pore's area. Averaged radial profiles show a peak of the intensity at
normalized radius $R_\mathrm{N} = r /R_\mathrm{pore} = 2.1$, followed by maxima
of the divergence at $R_\mathrm{N}= 2.3$ and the radial component of the
horizontal velocity at $R_\mathrm{N}= 4.6$. The divergence is negative within
pores strongly suggesting converging flows towards the center of pores, whereas
exterior flows are directed towards neighboring supergranular boundaries. The
photometric radius of pores, where the intensity reaches quiet-Sun levels at
$R_\mathrm{N} = 1.4$, corresponds to the position where the divergence is zero
at $R_\mathrm{N} = 1.6$.}
% conclusions heading (optional)
{Morphological and photometric properties as well as horizontal flow fields have
been obtained for a statistically meaningful sample of pores. This provides
critical boundary conditions for MHD simulations of magnetic flux
concentrations, which eventually evolve into sunspots or just simply erode and
fade away. Numerical models of pores (and sunspots) have to fit within these
confines, and more importantly ensembles of pores have to agree with the
frequency distributions of observed parameters.}


\maketitle

%###############################################################################
%#
%#    INTRODUCTION
%#
%###############################################################################

\section{Introduction}

% hierarchy of solar magnetic fields: flux tubes -> pores
In the hierarchy of solar magnetic fields individual flux
tubes, which may appear as bright points \citep[e.g.,][]{Steiner2001,
Schuessler2003}, are followed in size by magnetic knots and azimuth centers with
a similar size as pores but with no decrease in continuum intensity
\citep{Keppens1996}. These mark the transition to stable features (`pores'),
which have sufficient magnetic flux and a high filling factor of kilogauss flux
tubes to inhibit convective energy transport and to last for several hours and
up to days \citep[e.g.,][]{Suetterlin1998b}.

% hierarchy of solar magnetic fields: pores -> sunspots
Pores also demarcate the transitory state between small-scale magnetic elements
and sunspots. The transition from a pore to a sunspot is driven by changes of
the enclosed magnetic flux. According to \citet{Rucklidge1995}, there is an
overlap between large pores and small sunspots. The convective mode responsible
for this overlap sets in suddenly and rapidly, when the inclination to the
vertical of the photospheric magnetic field exceeds some critical value. The
consequence of this convective interchange is the filamentary penumbra, which
governs the energy transport across the boundary of the spot into the external
plasma. Numerical modeling of \citet{Tildesley2004} confirms the existence of
convectively driven filamentary instabilities, which lead to sunspot formation
-- starting with small pores gathering magnetic flux through magnetic pumping.
Reviewing the most important properties of pores (magnetic
field, temperature, Doppler velocity, and horizontal proper motions) sets the
stage for our statistical study of the dynamical environment around pores.

% magnetic field
Pores, often described as sunspots lacking a penumbra \citep{Bray1964}, form by
the advection of magnetic flux and clustering of magnetic elements. The magnetic
field of dark pores with typical diameters of 1.5--3.5~Mm is always in excess of
1500~G. \citet{Suetterlin1996} derive the magnetic field strength (about 1800~G
at the center) and inclination (about $70\degr$ at the periphery) as a function
of the distance from the pore's center. \citet{Keppens1996} present one of the
few studies with a statistically significant sample of 51 pores and 22 azimuth
centers in four active regions. At a critical magnetic flux of (4--5)\,$\times
10^{19}$~Mx, azimuth centers (magnetic filling factor always less than 50\%)
undergo a transformation and become pores with a higher magnetic filling factor
(25/65\% at their periphery/center). The magnetic flux linked to pores at their
periphery provides a mechanism for keeping them in their existing state and for
contributing to further growth. If magnetic flux from the surroundings is
constantly added to pores, then their growth can be maintained \citep{Wang1992}.
The magnetic radius of a pore is always larger than its
continuum radius, which points to the existence of a magnetic canopy
\citep{Suetterlin1996}.

% temperature
Image restoration holds the key to unravel the fine structure
of sunspots and pores \citep[e.g.,][]{Denker1998b}. Their temperature structure
can be assessed by three-color photometry \citep{Suetterlin1998a}. In both
sunspots and pores, umbral dots never reach photospheric temperatures. Umbral
dots exceed the about 2000~K cooler background by about 900--1300~K. Umbral
oscillations have been observed by \citet{Balthasar2000}, and
\citet{Sobotka1999} find (quasi-)oscillatory brightness variations of umbral
dots. In general, the intensity of a flux tube depends on the balance of
radiative heating along its perimeter and radiative cooling proportional to its
cross-section \citep[e.g.,][]{Suetterlin1998b}. In pores, the latter dominates,
thus a pore is dark.

% LOS velocities 
Flows play a significant role in development of pores.
\citet{Keil1999} observed in an active region with newly emerging magnetic flux
that pores form by an increased concentration of magnetic fields at the
supergranular boundaries and that surface motions intensify this process.
Downflows appear in annular, ring-like structures around most of the pores
\citep[e.g.,][]{Hirzberger2003, Sankarasubramanian2003}, and the line-of-sight
(LOS) component of the magnetic flux increases with the downflow velocity. The
strength of the LOS velocity rises as one moves down from the upper to the lower
photosphere. \citet{Uitenbroek2006} discovered a strong
supersonic downflow (more than 24~km~s$^{-1}$) in close proximity to a pore
using the Ca\,\textsc{ii} $\lambda$854.21~nm infrared triplet line. They
interpret the downflow as the signature of a siphon flow \citep{Thomas1993,
Montesinos1997} along a flux tube connecting two magnetic flux concentrations of
opposite polarity.

% horizontal proper motions
\citet{Roudier2002} state that horizontal plasma velocities are smaller by a
factor of two to three inside pores \citep[see also][]{Keil1999}. The highest
velocities occur near the pore's border, where they are contaminated by
convective flows penetrating from the outside. \citet{Hirzberger2003} finds that
the horizontal flow fields are asymmetric and that the absolute values of
horizontal flow velocities are much smaller than the velocities reported by
\citet{Roudier2002}. Furthermore, positive divergence structures surrounding
pores are indicative of horizontal inflows, which can be explained either in
terms of the continuous downflows detected in LOS velocities or by continuous
explosive events in the granulation around the pore. The latter explanation
agrees with the discovery of `rosettas' \citep{Sobotka1999} -- a typical
divergence pattern related to mesogranulation. \citet{Keil1999}
point out that the perturbation of a pore by (exploding) granules can initiate
the formation of a penumbra, which happens rapidly in just a few minutes
\citep{Leka1998, Yang2003a}. 

% moving magnetic features
The connection between the moat flow and moving magnetic
features (MMFs) around pores is still not clearly established. \citet{Deng2007}
notice the moat flow, which is a characteristic of sunspots, also around a
residual pore after a sunspot decayed and lost its penumbra.
\citet{Zuccarello2009} observe MMFs and moat flow in the vicinity of a naked
sunspot, i.e., even when penumbral filaments and the Evershed flow are not
present \citep[cf.,][]{CabreraSolana2006}. Similarly, \citet{Verma2012a} detect
MMFs and moat flow around a spot with almost no filamentary penumbra. In
contrast, studying the flow fields around seven pores,
\citet{VargasDominguez2010} find no trace of moat flow. They relate inward and
outward motion around pores to exploding granules. Examining the same naked
sunspot as \citet{Zuccarello2009}, \citet{Dalda2012} conclude that MMFs can be
explained in the same way as for sunspots with penumbral filaments, because of
the same underlying magnetic structure. 

% Table 1
\begin{table}[t]
\caption{Time-series of G-band images suitable for LCT.}\label{TAB01}
\begin{center}
\footnotesize
\begin{minipage}{60mm}
\begin{tabular}{cccc}
\hline\hline
Active region & Date & $N_s$& $N_p$\rule[-2mm]{0mm}{6mm}\\
\hline
NOAA~10921 & 2006-11-03  &    &     \rule[0mm]{0mm}{4mm}\\
           & 2006-11-05  & 11 & 146 \\
NOAA~10926 & 2006-11-30  &    &     \\
           & 2006-12-04  & 27 & 69  \\
NOAA~10930 & 2006-12-07  & 30 & 242 \\
NOAA~10933 & 2007-01-10  & \phn6 & \phn29\\
NOAA~10938 & 2007-01-15  &    & \\
           & 2007-01-20  & 53 & 287\\
NOAA~10940 & 2007-02-01  &    & \\
           & 2007-02-05  & 40 & 269\\
no region No. & 2007-03-08 & \phn3 & \phnn3\\
NOAA~10953 & 2007-04-27  &    & \\
           & 2007-05-06  & 79 & 315\\
NOAA~10956 & 2007-05-16  &    &        \\
           & 2007-05-22  & 10 & \phn51 \\
NOAA~10960 & 2007-06-04  &       &       \\
           & 2007-06-12  & \phn5 & \phn14\\
NOAA~10969 & 2007-08-24  &       &       \\
           & 2007-08-31  & \phn6 & \phn11\\
NOAA~10978 & 2007-12-07  &    &        \\
           & 2007-12-15  & 81 &  1421  \\
NOAA~10978 & 2008-01-06  & \phn5 & \phnn6\\
\hline
\textbf{Total} &       & \textbf{356} &  \textbf{2863}\rule[1mm]{0mm}{2mm}\\
\hline
\end{tabular}
\end{minipage}
\tablefoot{The number of one-hour sequence $N_s$ and the number of pores $N_p$
identified in the FOV are given for each active region.}
\end{center}
\end{table}

% summary - case vs. statistical studies
In summary, previous research efforts focus for the most
part on individual pores. Our statistical analysis has the goal to
comprehensively describe pores in terms of their morphological and photometric
properties and to characterize the associated flow fields.  This
study continues the series of articles \citep{Verma2011, Verma2012a, Verma2012b,
Verma2013}, where LCT has been carefully evaluated and established as an apt and
versatile tool to investigate horizontal proper motions in active regions,
sunspots, and pores.

% Data from space missions offer the opportunity for this type of research, 
% because of the uniform data quality and the absence of seeing, which allows us 
% to directly compare pores in different environments and at various stages of 
% their evolution.

%###############################################################################
%#
%#    OBSERVATIONS
%#
%###############################################################################

\section{Observations}

% FIGURE 1
\begin{figure*}[t]
\includegraphics[width=\columnwidth]{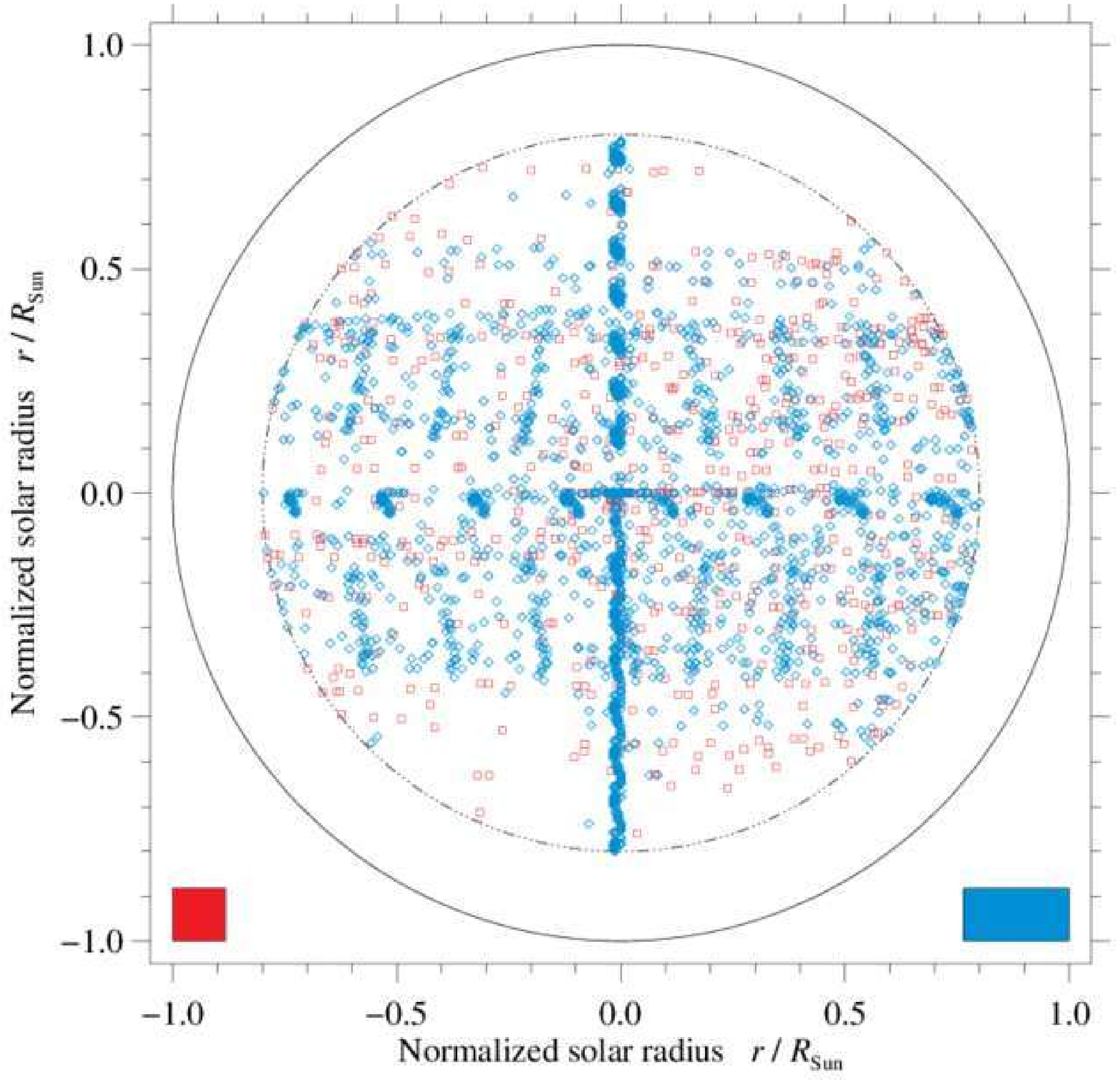}
\includegraphics[width=\columnwidth]{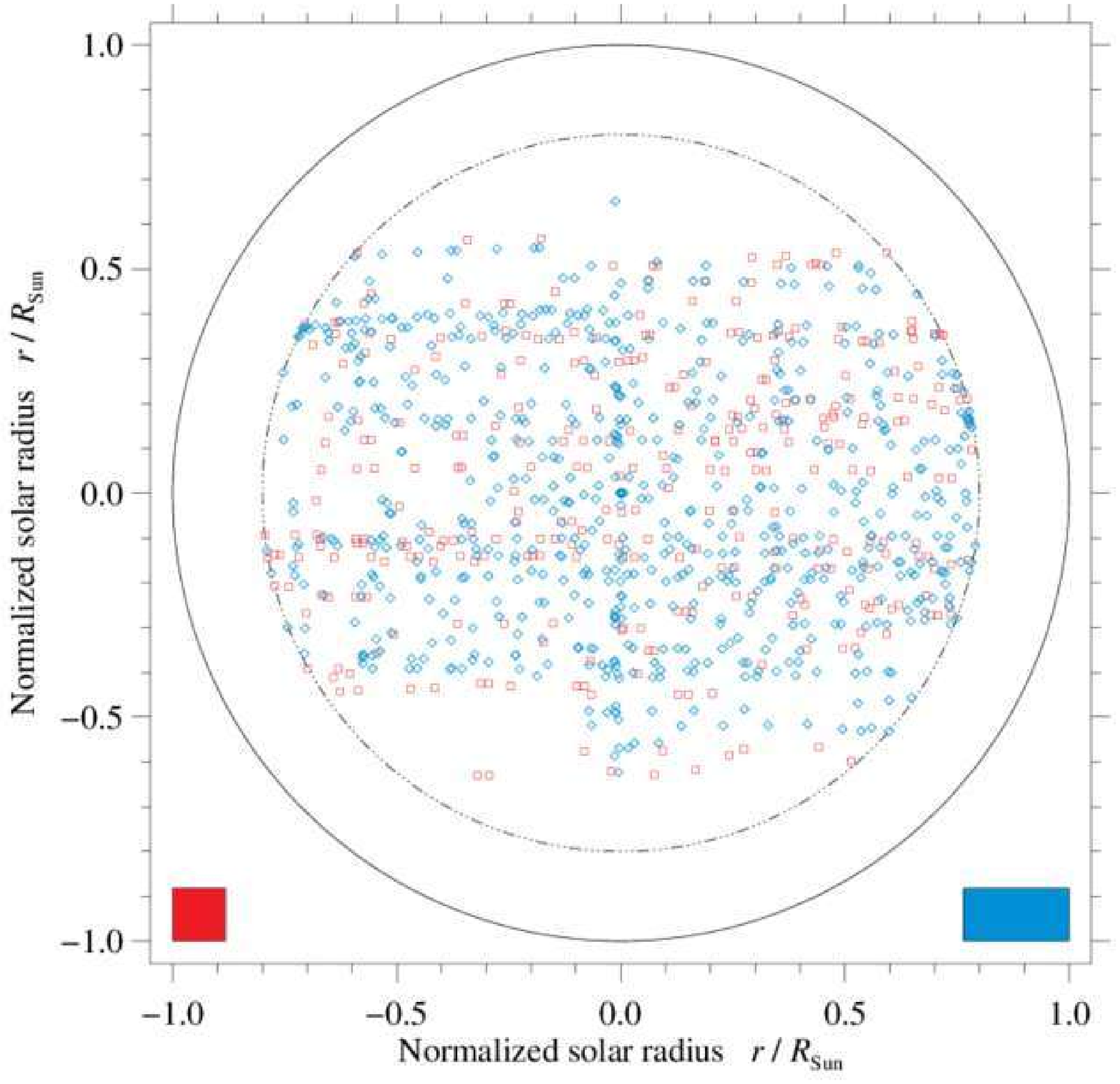}
\caption{Distribution of all 4530 selected single G-band images on the solar
disk in the period from 2006 October 25 to 2013 August~31 (left). Diamonds and
squares indicate single G-band images with $2048 \times 1024$ (3819 images) and
$1024 \times 1024$ (711 images) pixels, respectively. Distribution of all 1013
G-band images (690 images with $2048 \times 1024$ and 323 images with $1024
\times 1024$ pixels) with pores on the solar disk (right). The different image
sizes are indicated by the filled boxes in the lower left and right corners. The
solid line encircles the solar disk, and the dash-dotted line refers to $\mu =
\cos\theta = 0.6$.}
\label{FIG01}
\end{figure*}

G-band images (bandhead of the CH molecule at $\lambda
430.5$~nm) are routinely obtained with the broad-band filter imager (BFI) of
Hinode SOT \citep{Tsuneta2008, Kosugi2007}. Exploiting the high-contrast and
rich structural details of G-band images, \citet{Verma2011} have created a
database of flow maps, which is the foundation of this study. In the following,
we briefly recapitulate the particulars of data selection and analysis steps.
The initial data selection criterion requires at least a one-hour sequence with
a time cadence better than 100~s. The paucity of full-resolution data limits us
to images with only half the spatial resolution (0.11\arcsec\ pixel$^{-1}$) and
2$\times$2-pixel binning. For the time period from November 2006 to January
2008, we find about 153 data sets with $1024 \times 1024$ pixels
($111.6\arcsec \times 111.6\arcsec$) and 48 with $2048 \times
1024$ pixels ($223.2\arcsec \times 111.6\arcsec$). These
fields-of-view (FOV) are sufficiently large to cover significant parts of active
regions. The typical time cadence of the image sequences is 30--90~s. These
data sets have been split in 60-minute sequences with 30-minute overlap between
consecutive sequences. In total, we have about 557 one-hour sequences covering
various scenes on the solar surface, out of which 410 contained active regions.
However, the coverage is limited to a set of twelve
active regions (Table~\ref{TAB01}). Even though observing each region lasts
sometimes several days, a selection bias enters into the statistical study,
which is alleviated to some extend by covering pores at various stages of
evolution. In any case, the present study is the first attempt to use the
Hinode database to establish a statistically meaningful characterization of the
horizontal flow fields associated with solar pores. 

% FIGURE 2
\begin{figure*}[t]
\includegraphics[width=\textwidth]{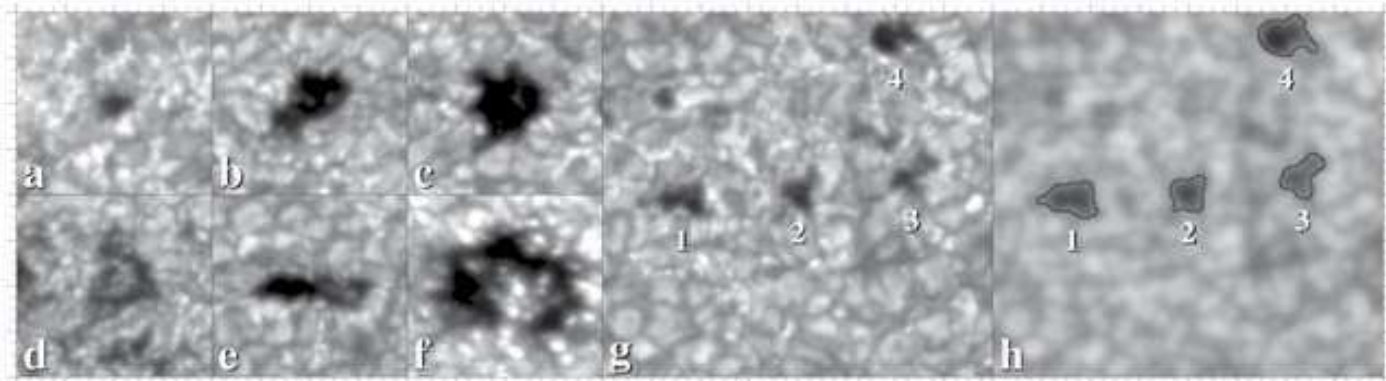}
\caption{Sample of different types of pores (Panels~a--g) extracted from the
    full-sized G-band images. The images are displayed in the range
    0.4--1.4\,$I_0$. Major tick marks are separated by 2~Mm. The application of
    a Perona-Malik filter to the G-band image in Panel~g is shown in Panel~h.
    This filter facilitates the extraction pores by intensity thresholding. The
    contours of the pores are marked by solid lines in Panel~h. Characteristic
    parameters of the pores are given in Table~\ref{TAB02}.}
\label{FIG02}
\end{figure*}

% TABLE 2
\begin{table*}[t]
\caption{Parameters describing the pores (a--g$_n$) depicted in
    Fig.~\ref{FIG02}.}
\normalsize\small
\begin{center}
\begin{tabular}{ccccccccccc}
\hline\hline
Pore & Date & Time & E--W & N--S & $\mu$ & $A_\mathrm{pore}$ & $l_\mathrm{pore}$
& $I_\mathrm{mean}$ & $I_\mathrm{min}$ & $\varepsilon$\rule[-2mm]{0mm}{6mm}\\
\hline
a\phs  & 2011-05-02 & 08:17~UT & \phn         $ 364.3\arcsec$ &     
$-258.1\arcsec$ & 0.88 & \phn 1.4~Mm$^2$ & \phn 4.6~Mm & 0.68 & 0.51 &
0.63\rule{0mm}{4mm}\\
b\phs  & 2010-03-14 & 10:32~UT &              $-175.2\arcsec$ & \phn $
378.7\arcsec$ & 0.90 & \phn 7.1~Mm$^2$ &     11.9~Mm & 0.56 & 0.31 & 0.83\\
c\phs  & 2011-06-02 & 18:04~UT & \phn         $ -36.7\arcsec$ &     
$-380.6\arcsec$ & 0.91 & \phn 9.9~Mm$^2$ &     14.3~Mm & 0.53 & 0.28 & 0.53\\
d\phs  & 2007-01-18 & 20:48~UT & \phn\phn\phn $   7.5\arcsec$ & \phn $
136.3\arcsec$ & 0.99 & \phn 8.9~Mm$^2$ &     17.0~Mm & 0.71 & 0.51 & 0.84\\
e\phs  & 2012-05-03 & 15:42~UT & \phn\        $ -31.4\arcsec$ &     
$-209.1\arcsec$ & 0.97 & \phn 7.3~Mm$^2$ &     13.9~Mm & 0.63 & 0.33 & 0.93\\
f\phs  & 2011-06-04 & 15:40~UT & \phn         $ 625.3\arcsec$ & \phn $
266.9\arcsec$ & 0.69 &     21.5~Mm$^2$ &     29.2~Mm & 0.60 & 0.34 & 0.74\\
g$_1$  & 2007-05-12 & 10:56~UT & \phn         $ 247.1\arcsec$ & \phn $
-98.0\arcsec$ & 0.96 & \phn 2.5~Mm$^2$ & \phn 6.9~Mm & 0.66 & 0.46 & 0.77\\
g$_2$  & 2007-05-12 & 10:56~UT & \phn         $ 253.4\arcsec$ & \phn $
-97.8\arcsec$ & 0.96 & \phn 1.7~Mm$^2$ & \phn 5.5~Mm & 0.67 & 0.47 & 0.49\\
g$_3$  & 2007-05-12 & 10:56~UT & \phn         $ 259.6\arcsec$ & \phn $
-96.8\arcsec$ & 0.96 & \phn 1.8~Mm$^2$ & \phn 6.3~Mm & 0.71 & 0.52 & 0.76\\
g$_4$  & 2007-05-12 & 10:56~UT & \phn         $ 258.7\arcsec$ & \phn $
-88.5\arcsec$ & 0.96 & \phn 2.7~Mm$^2$ & \phn 7.2~Mm & 0.64 & 0.43 &
0.67\rule[-2mm]{0mm}{4mm}\\
\hline
\end{tabular}
\tablefoot{The heliographic coordinates of the pores are given along
    with $\mu = \cos\theta$. The area 
    $A_\mathrm{pore}$ and perimeter length $l_\mathrm{pore}$ are based on the
    pixels belonging to the pore, which were used to compute the mean
    $I_\mathrm{mean}$ and minimum $I_\mathrm{min}$ intensities. Ellipse fitting
    to the enclosed pixels yields the numerical eccentricity $\varepsilon$.}
\end{center}
\label{TAB02}
\end{table*}

Basic image-processing steps in preparation for LCT
include subtraction of dark current, gain calibration, removal of spikes caused
by high-energy particles, correction of geometrical foreshortening, resampling
onto a regular grid of 80~km $\times$ 80~km, removal of the center-to-limb
variation (CLV), and intensity normalization such that the quiet-Sun intensity $I_0$
corresponds to unity. Images have been aligned with respect to the first image
in a sequence by applying the shifts between consecutive images in succession
using cubic spline interpolation with subpixel accuracy. These shifts are
calculated using the cross-correlation over the central part of the images. The
signature of the five-minute oscillation has been removed from the images by
using a three-dimensional Fourier filter with a cut-off velocity of
8~km~s$^{-1}$ corresponding roughly to the photospheric sound speed. A high-pass
filter in form of a Gaussian kernel with a FWHM of 15 pixels (1200~km) has been
applied to all images to suppress strong intensity gradients. The LCT algorithm
of \citet{Verma2011} builds upon the seminal work of \citet{November1988}. The
algorithm operates on subimages with a size of $32 \times 32$ pixels
corresponding to 2560~km $\times$ 2560~km on the solar surface. A Gaussian
similar to the kernel used in the high-pass filter serves as an
apodization/sampling window limiting the cross-correlation to structures with a
size of 1200~km, i.e., roughly the size of a granule.
Locating the maximum of the cross-correlation function
yields a displacement vector, and dividing by the time interval between
successive images leads to a velocity vector. One-hour averages are computed for
each pixel, and the resulting flow maps are the basis for this study.

In principle, time-averaged G-band images, which
belong to the data set above, are sufficient to derive the morphological and
photometric properties of pores. However, including `single' G-band images into
the morphological and photometric study, which are not restricted by LCT
requirements, significantly broadens the scope and scientific reach of the
study. In addition, comparing time-averaged with instantaneous results gives a
better understanding of how accurate the measured statistical properties are.
Most importantly, now the entire database of Hinode G-band images is at our
disposal, which covers more than half a solar cycle.

We restrict our selection of single G-band images to locations
on the solar disk, where the center of the FOV is within a circle corresponding
to $\mu = \cos \theta = 0.6$, where $\theta$ is the heliocentric angle, i.e.,
images in close proximity to the solar limb are excluded. On each day, we select
one G-band image for each of the observed targets. Targets are considered to be
different, if the telescope pointing on the solar disk differ by more than $0.1
\mu R_\sun$, where $R_\sun$ is the solar radius. Thus, continuous tracking of an
active region over the course of a day can produce two or three images of the
same (active) region. The above selection criterion guarantees that images of
the same region are spread over time (about 10~hours), so that pores will have
significantly evolved, thus reducing potential biases.

The single G-band images cover the time period from 2006
October~25 to 2013 August~31, which includes the deep minimum of solar cycle
No.~23 and the rise towards the maximum of cycle No.~24. In total 4530 G-band
images have been downloaded. However, only 1013 G-band images contain pores,
whereas the other images are related to quiet-Sun studies -- in particular
studies of the center-to-limb variation. The disk coverage of the initial
selection of G-band images is indicated in Fig.~\ref{FIG01} for both image
sizes. Specifically, the $2048 \times 1024$-pixel images exhibit a peculiar
pattern, which consists of two parts: (1) a cross-shaped pattern along the
central meridian and the equator, which is typical for center-to-limb variation
studies of the quiet Sun and (2) a pattern tracing certain sections of great
circles within the solar activity belts. The emergent pattern bears close
resemblance to a bright reflection (`the Pope's revenge'), which appears when
the Sun illuminates the stainless steel dome of the Berlin television tower at
Alexanderplatz.

This statistical study is based only on a photometric and
morphological analysis of G-band images and LCT flow maps, i.e., neither
measurements of the magnetic field nor any type of spectroscopic observations
are included. In addition, even though most Hinode G-band observations are
time-series, we do not include any dynamical aspects related to the evolution of
pores. Consequently, the scientific focus is narrower, which however is
compensated by a much improved statistical sample of many thousands of solar
pores.

%###############################################################################
%#
%#    FEATURE IDENTIFICATION AND SELECTION
%#
%###############################################################################

\section{Feature identification and selection}

% FIGURE 3
\begin{figure}[t]
\includegraphics[width=\columnwidth]{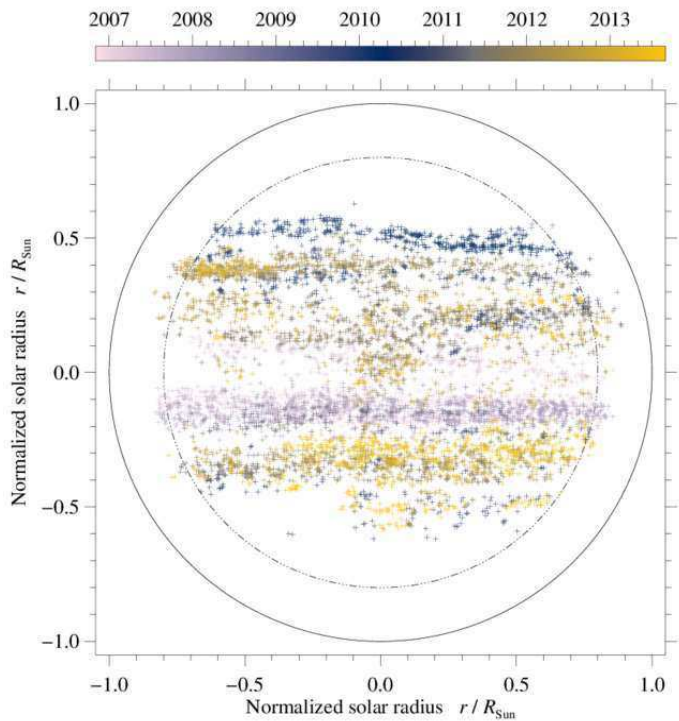}
\caption{Disk-coverage of all 7643 pores detected in single G-band images. The
    scale indicates the date when the pores were observed.}
\label{FIG03}
\end{figure}

The identification and selection process of pores follows
the same scheme for both single and time-averaged G-band images. Nevertheless,
examples discussed in this section are based on pores in single G-band images.
A pore is dark feature on the solar surface lacking a penumbra
\citep[e.g.,][]{Suetterlin1996}. The magnetic field is already sufficiently
strong to suppress the convective energy transport. This physical property leads
to the first selection criterion: at least one dark core ($I_\mathrm{core} <
0.6\,I_0$) of at least 10 contiguous pixels must be enclosed by the perimeter of
the pore. Additionally, two size thresholds are used to bracket pores: (1) dark
features with 125 pixels (0.8~Mm$^2$) or are classified more appropriately as
magnetic knots, and (2) dark structures with more than \mbox{15\;\!625} pixels
(100~Mm$^2$) are in all likelihood sunspots, even though there is a substantial
overlap in size between sunspots and pores \citep[e.g.,][]{Rucklidge1995}.

The criterion that a pore must possess a dark core is very
strict and excludes two types of small-scale dark features: (1) darkened patches
(intensities down to about 0.8\,$I_0$) with a size of one to three granules
within the quiet-Sun granulation and (2) roundish dark features (intensities
down to about 0.6\,$I_0$) with the size of a granule and below within plage
regions related to `old' magnetic flux. The first type corresponds likely to
azimuth centers \citep{Keppens1996, Leka1998, Leka2001}, whereas the latter
coincides with either flux concentrations due to flux pile-up at supergranular
boundaries \citep{Keil1999} or with remnants of decaying pores. This limitation
can be alleviated to some extend by correcting instrument stray-light and the
telescope's modulation transfer function \citep[see][]{Mathew2009}. However, the
iterative maximum-likelihood deconvolution algorithm is computationally
expensive, in particular when many thousands of large-format G-band images are
involved. Therefore, we have opted against this type of image correction.

% FIGURE 4
\begin{figure}[t]
\includegraphics[width=\columnwidth]{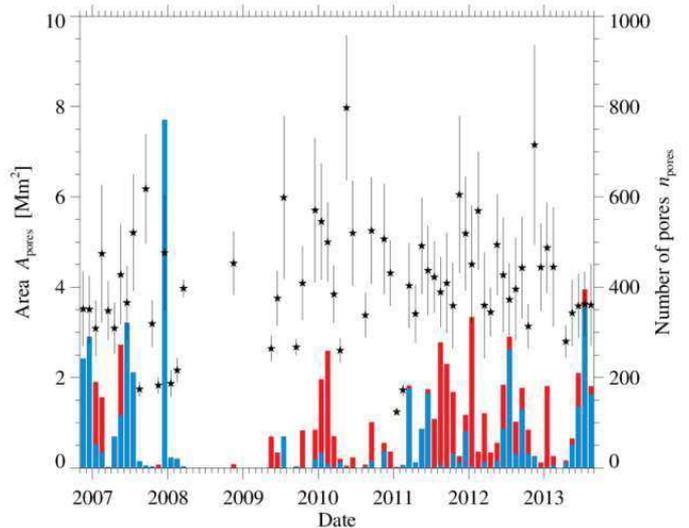}
\caption{Mean area of pores (`$\scriptstyle\bigstar$' symbols) for each month 
    derived from single G-band images. The vertical lines represent
    \nicefrac{1}{5}$^\mathrm{th}$ of the standard deviation of the monthly
    distributions. The red and blue bars correspond to the number of pores per
    month (\emph{right scale}) in the northern and southern hemisphere,
    respectively.}
\label{FIG04}
\end{figure}

Simple intensity thresholding is too restrictive in determining
the boundaries of a pore and results in very corrugated boundaries.
\citet{Perona1990} have suggested an approach to detect `semantically
meaningful' edges at coarser spatial scales using anisotropic diffusion. While
Gaussian smoothing just blurs the image, the Perona-Malik filter conserves the
boundaries of an object. We select a conduction coefficient for the anisotropic
diffusion equation which favors wider over smaller regions. The property of
intra-region smoothing is very powerful in detecting contiguous dark features by
thresholding the filtered gray-scale image. Implementing the diffusion process
results in an iterative numerical scheme. The process is stopped after 50
iterations, and an intensity threshold of $0.85\,I_0$ is applied to the filtered
images identifying features resembling pores (see Fig.~\ref{FIG02}h). A visual
inspection is still required to confirm that the features are actually pores.
Various properties of pores can now be computed for all pores from the
unfiltered images using the contours of the filtered images as described below.
Table~\ref{TAB02} summarizes the results for the sample of pores shown in
Fig.~\ref{FIG02}. Processing more than 4500 single G-band images takes about
about 1350~min on an Intel Xeon X5460 CPU at 3.16~GHz.

The morphological differences between pores are significant
(see Fig.~\ref{FIG02}). Many pores contain umbral dots \citep{Sobotka1999} and
faint (sometimes even strong) light-bridges. Pores can exist in isolation or in
proximity to sunspots or other pores. Some pores form chains and even ring-like
structures. All these characteristics are not criteria in the visual selection
process. The only strong discriminator is the presence of any type of elongated
feature resembling penumbral filaments. These are taken as indications that the
magnetic field is no longer close to vertical. Furthermore, once a penumbra
starts to form, the Evershed flow sets in and the interaction of the magnetic
fields with the surrounding plasma has been significantly altered
\citep[e.g.,][]{Leka1998, Yang2003a, Langhans2005}. Therefore, any type of
filamentary structure within the confines of the identified feature but also in
its close proximity results in its exclusion.

% FIGURE 5
\begin{figure*}[t]
\includegraphics[width=0.5\textwidth]{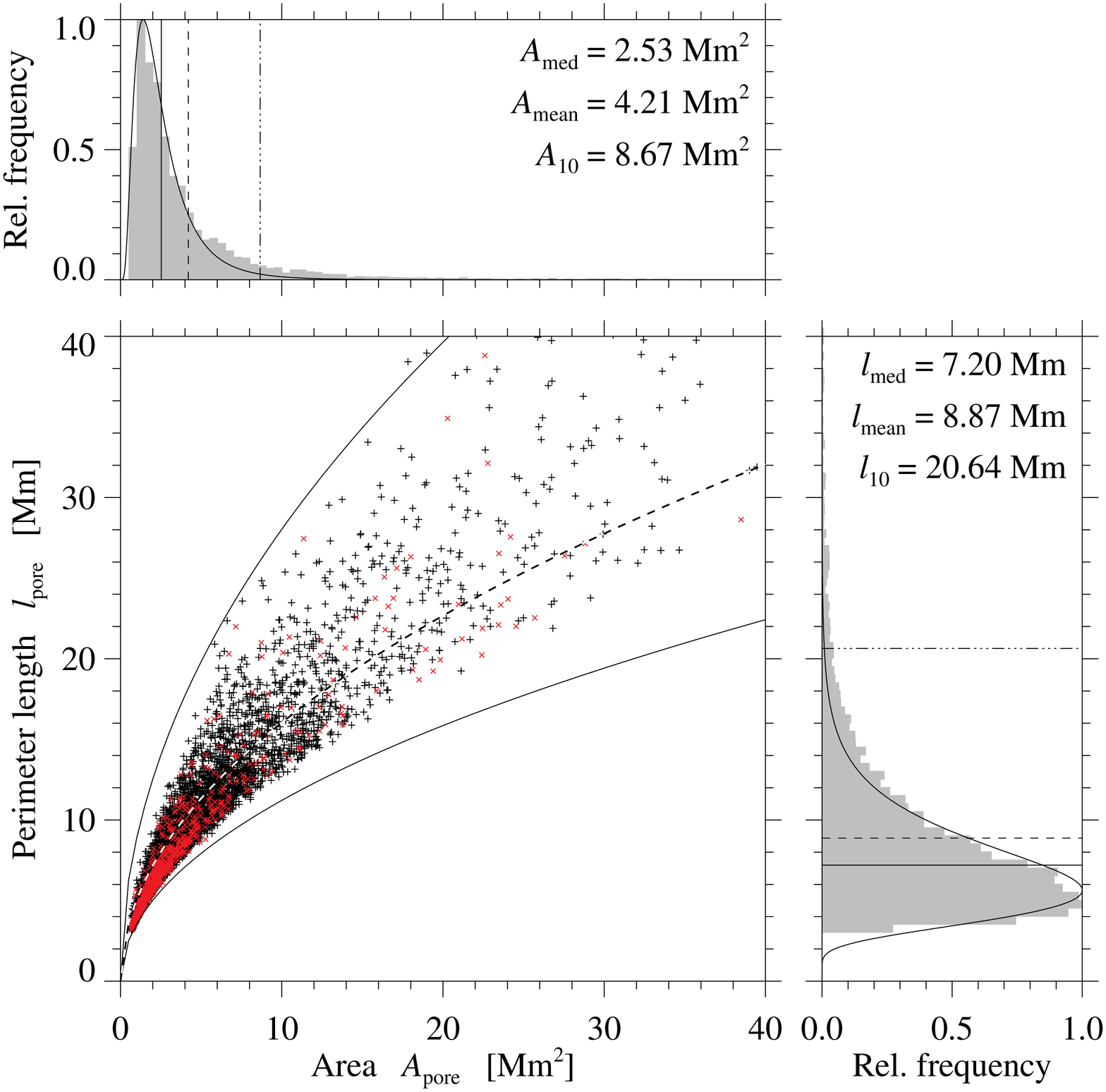}
\includegraphics[width=0.5\textwidth]{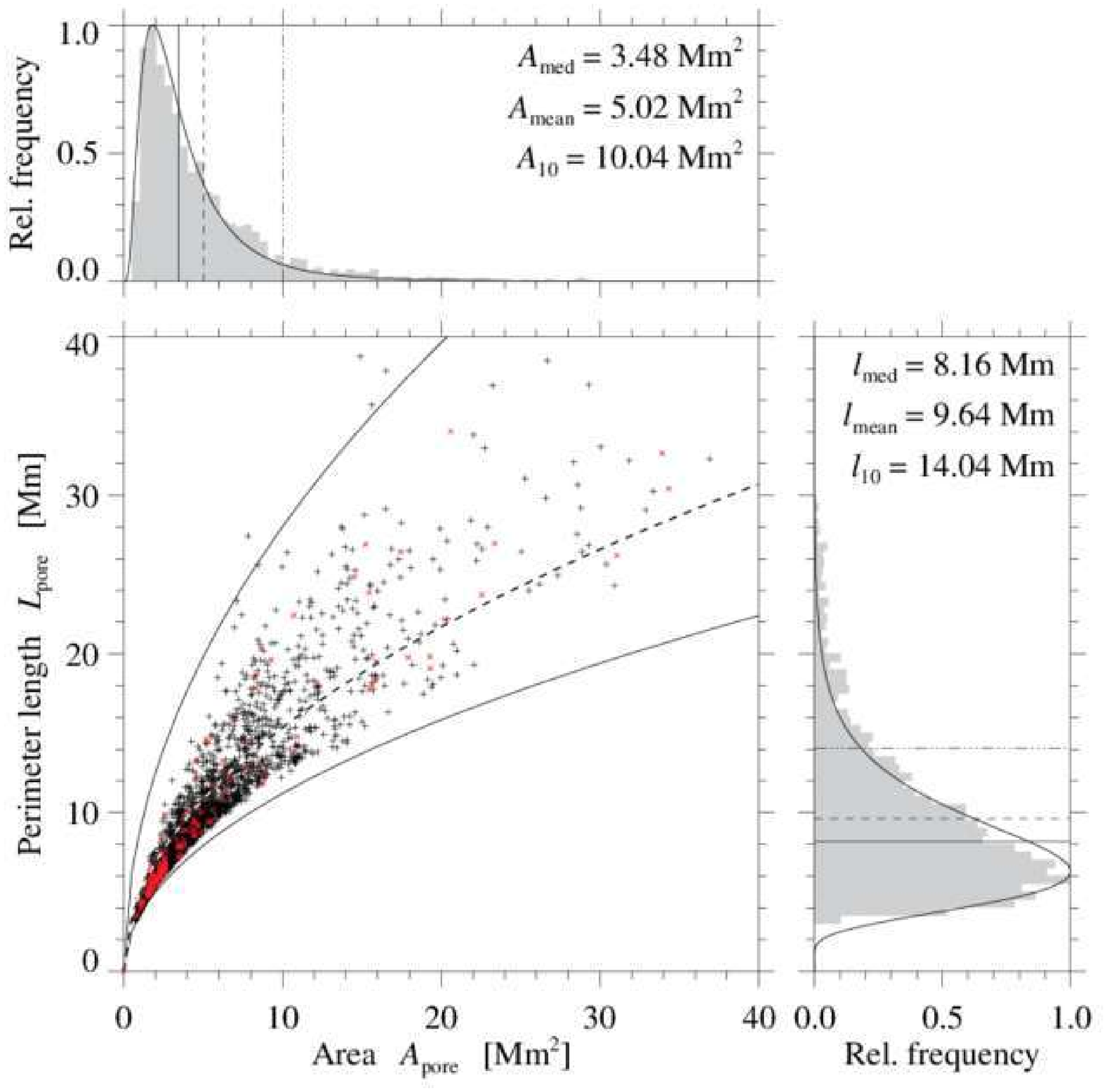}
\caption{Scatter plots of perimeter length
    $l_\mathrm{pore}$ vs.\ area $A_\mathrm{pore}$ for single (\textit{left})
    and time-averaged (\textit{right}) G-band images, which also include the
    respective frequency distributions. Red symbols `$\scriptstyle\times$' mark
    isolated pores. The lower envelope \mbox{(\emph{solid}\tsp)} of the scatter
    plots refers to perfectly circular pores ($\alpha = 1$ in Eqn.~\ref{EQN01}),
    while the upper envelope \mbox{(\emph{solid}\tsp)} is given by a scaling
    factor of $\alpha = 2.5$. Non-linear least squares fits
    \mbox{(\emph{dashed}\tsp)} using Eqn.~\ref{EQN01} lead to scaling factors of
    $\alpha = 1.43$ and $1.37$ for single and time-averaged G-band images,
    respectively. Median, mean, and $10^\mathrm{th}$ percentile of the frequency
    distributions (\emph{light gray bars}) are given by solid, dashed, and
    dash-dotted lines, respectively. Quantitative values of the measures are
    specified in the panels for perimeter lengths $l_\mathrm{pore}$ and areas
    $A_\mathrm{pore}$. Log-normal frequency distributions
   \mbox{(\emph{solid}\tsp)} properly fit the histograms.}
\label{FIG05}
\end{figure*}

In the visual inspection, 438 features are classified as orphan
penumbrae, where the filamentary structure dominates the dark cores, 1511
features are sunspots with rudimentary penumbrae, and 68 features are
categorized as sunspots because their penumbra encompasses more than half of the
sunspot. In 434 cases, features are too close to the boundary of the G-band
images and have to be excluded from further analysis. In total, almost
\mbox{10\;\!000} features have been inspected and 7643 pores are kept for
further analysis. Yet another 2863 pores have been
identified in time-averaged G-band images (Table~\ref{TAB01}). Thus, to the best
of our knowledge, these two data sets comprise the largest sample of pores used
in any type of statistical study so far.

% \subsection{Parameters describing morphology of pores and associated flow
% fields}

% TABLE 3
\begin{table}
\caption{Scaling factors $\alpha$ and $\alpha_\mathrm{fit}$ of the square-root
dependence between perimeter length $l_\mathrm{pore}$ and area $A_\mathrm{pore}$
of a pore.}
\begin{center}
\begin{tabular}{ccccccc}
\hline\hline
& \multicolumn{6}{c}{{\rule{9mm}{0mm}Area interval
[Mm$^2$]\rule{9mm}{0mm}}}
\rule[-1mm]{0mm}{5mm}\\
\cline{3-6}
                      & 0--5 & 5--10 & 10--20 & 20--30 & 30--40 & 
\textbf{0--80} \rule[-1mm]{0mm}{4.5mm}\\
\hline
\hline
& \multicolumn{6}{c}{{\rule{9mm}{0mm}Single G-band
images\rule{9mm}{0mm}}}\rule[-1mm]{0mm}{5mm}\\
\cline{3-6}
$\alpha$              & 1.28 & 1.42 & 1.51 & 1.66 & 1.85 & \textbf{1.43}
\rule{0mm}{3.5mm}\\
$\alpha_\mathrm{fit}$ & 1.06 & 1.06 & 1.11 & 1.11 & 1.11 & \textbf{1.06}
\rule{0mm}{3.5mm}\\
\hline
& \multicolumn{6}{c}{{\rule{9mm}{0mm}Time-averaged G-band
images\rule{9mm}{0mm}}}\rule[-1mm]{0mm}{5mm}\\
\cline{3-6}
$\alpha$              & 1.23 &  1.38 &   1.48  &  1.59 &   1.54 &
\textbf{1.38} \rule{0mm}{3.5mm}\\
$\alpha_\mathrm{fit}$ & 1.05 &  1.05 &   1.05 &   1.06 &   1.11 &
\textbf{1.05} \rule[-1mm]{0mm}{3mm}\\
\hline
\end{tabular}
\end{center}
\label{TAB03}
\end{table}

The feature identification leads to binary masks containing all
pixels belonging to pores. The masks provide easy access to other physical
quantities such as the normalized intensity, the position on the solar disk in
heliocentric coordinates, and the cosine of the heliocentric angle $\mu$ (see
Table~\ref{TAB02}). Standard tools for `blob analysis' \citep{Fanning2011} are
used to derive parameters describing the morphology and
the associated flow field of pores. We measure the area
covered by pores $A_\mathrm{pore}$, the length of the circumference
$l_\mathrm{pore}$, the mean intensity $I_\mathrm{mean}$, the average horizontal
flow speed $v_\mathrm{mean}$, the divergence \mbox{$\nabla\ \cdot v$}, and the
vorticity $\nabla \times v$ within pores. The most important
ones are the perimeter length (circumference) $l_\mathrm{pore}$ and area
$A_\mathrm{pore}$. The connectivity of pixels inside a pore is 4-adjacent, i.e.,
only horizontal and vertical neighbors are included but not diagonal neighbors.
A `chain code' algorithm \citep{Russ2011} computes the length and area from the
perimeter points. The thus calculated area is always smaller than the area
represented by the pixels masking the pore. Fitting an ellipse
to each pore yields among other parameters the semimajor $a$ and semiminor $b$
axes.
% The angle of the major axis with the horizontal $\theta$ corresponds to
% lines of equal latitude because of the chosen deprojection method.

%###############################################################################
%#
%#    RESULTS
%#
%###############################################################################

\section{Results}

The photometric results are based both on single and
time-averaged G-band images, whereas the solar cycle dependence is discussed
using the former data set, and the analysis of flow fields naturally relies on
the latter collection of pores. In the statistical analysis numerical errors are
typically small but systematic errors introduced by selection criteria and
preconceptions can strongly affect the outcome of a study. Using both single
and time-averaged G-band images to measure the photometric and morphological
properties of pores provides guidance in interpreting the results and ensures
that our conclusions are sound. 

\subsection{Solar cycle dependence}

The distribution of all pores identified on the solar disk in single G-band
images is presented in Fig.~\ref{FIG03}.
The color-coded `$\scriptstyle+$' symbols indicate the time and location of the
detection. Some pores are situated outside the circle with $\mu = 0.6$ because
the FOV covered by the G-band images extends beyond this circle, especially for
images with $2048 \times 1024$ pixels. We observe a pronounced north-south
asymmetry during the extended minimum of solar cycle 23 with a larger amount of
pores at low latitudes in the southern hemisphere. This trend changes, when most
of the pores of the new solar cycle 24 appear at high latitudes in the northern
hemisphere. This striking imbalance persists during the rise of solar cycle 24.
These observations are in good agreement with photospheric magnetic field
measurements at the National Solar Observatory/Kitt Peak \citep{Petrie2012}, in
particular with the vector spectro-magnetograph of the Synoptic Optical
Long-term Investigations of the Sun \citep[SOLIS,][]{Keller2003, Henney2009}
instrument suite. In general, the color-coded locations of the pores are an
easily perceptible graphic representation of Sp\"orer's law \citep[see the
introduction of][]{Leighton1969}.

Figure~\ref{FIG04} provides another view of the north-south
asymmetry of the pores' location on the solar disk. The height of the histogram
bars corresponds to the number of pores observed in each hemisphere per month.
Red and blue colors indicate the northern and southern hemisphere, respectively.
Most pores appear in the southern hemisphere during the decline and deep minimum
of solar cycle 23. Only in three months, more pores surface in the northern
hemisphere. The deep solar minimum also leaves its signature in the pore count.
Pores are virtually absent in Hinode G-band images for more than one year (March
2008 -- April 2009). The dearth of pores ceased at last in early 2010. The
north-south asymmetry reversed with the rise of solar cycle 24. However, there
are periods (most prominently in 2012 and 2013) when pores are more common in
the south. Interestingly, months with a clear hemispheric preference occur more
often than months with a balanced pore count. In addition, less pores are
detected per month in solar cycle 24 as compared to the previous cycle.

% TABLE 4
\begin{table}
\caption{Characteristics of the log-normal frequency distributions for
    area $A_\mathrm{pore}$ [Mm$^2$] and perimeter length 
    $l_\mathrm{pore}$ [Mm].}
\normalsize\small
\begin{center}
\begin{tabular}{lcccc}
\hline\hline
 & \multicolumn{2}{c}{Single}        & \multicolumn{2}{c}{Time-averaged}
\rule{0mm}{4mm} \\
 & \multicolumn{2}{c}{G-band images} & \multicolumn{2}{c}{G-band images}
\rule[-2mm]{0mm}{4mm} \\
\cline{2-5}
Variable $\ln x$ & $A_\mathrm{pore}$ & $l_\mathrm{pore}$ & $A_\mathrm{pore}$ &
$l_\mathrm{pore}$\rule[-2mm]{0mm}{6mm}\\
\hline
Mean $\ln x$, $\mu$                                      & \phn 0.77 & \phn 
1.91 & \phn  1.13 & \phn  2.03\rule{0mm}{4mm}\\
Standard deviation $\ln x$, $\sigma$                     & \phn 0.66 & \phn 
0.42 & \phn  0.73 & \phn  0.43\rule{0mm}{4mm}\\
Mean $e^{\mu + \sigma^2/2}$                              & \phn 2.69 & \phn 
7.37 & \phn  4.03 & \phn  8.37\rule{0mm}{4mm}\\
Median $e^\mu$                                           & \phn 2.17 & \phn 
6.74 & \phn  3.09 & \phn  7.62\rule{0mm}{4mm}\\
Mode $e^{\mu - \sigma^2}$                                & \phn 1.41 & \phn 
5.64 & \phn  1.82 & \phn  6.31\rule{0mm}{4mm}\\
10$^\mathrm{th}$  percentile                             & \phn 4.99 &     
11.53 & \phn  7.76 &      13.22\rule{0mm}{4mm}\\
Variance $(e^{\sigma^2}-1)\,e^{2\mu+\sigma^2}$           & \phn 3.88 &     
10.61 &      11.38 &      14.55\rule{0mm}{4mm}\\
Skewness $(e^{\sigma^2}+2)\sqrt{e^{\sigma^2}-1}$         & \phn 2.60 & \phn 
1.56 & \phn  3.10 & \phn  1.69\rule{0mm}{4mm}\\
Kurtosis $e^{4\sigma^2}\!\!+\!2e^{3\sigma^2}\!\!+\!3e^{2\sigma^2}\!\!-\!6$ &
                                                               13.98 & \phn 
3.74 &      20.86 & \phn  4.03\rule[-2mm]{0mm}{6mm}\\
\hline
\end{tabular}
\end{center}
\label{TAB04}
\end{table}

The large number of pores in the sample allows us to search for
a potential solar cycle dependence of physical parameters, e.g., the area of the
pores. Asterisks in Fig.~\ref{FIG04} refer to the average area of the pores for
a given month. The standard deviation for each monthly sample is large so that
the corresponding vertical bars have to be scaled down by a factor of five. The
average area of a pore for the entire sample is 4.21~Mm$^2$. The corresponding
values for solar cycles 23 and 24 are 4.12~Mm$^2$ and 4.25~Mm$^2$, respectively.
There is no obvious trend in area related to the rise or decline of the activity
cycle. The sample of pores presented in this study is certainly not complete.
This can only be accomplished with synoptic full-disk observations, which are
nowadays provided by the Helioseimic and Magnetic Imager
\citep[HMI,][]{Scherrer2012} of the Solar Dynamics Observatory
\citep[SDO,][]{Pesnell2012}. However, the reduced spatial resolution of HMI does
not permit the detection of small pores.

% FIGURE 6
\begin{figure}
\includegraphics[width=\columnwidth]{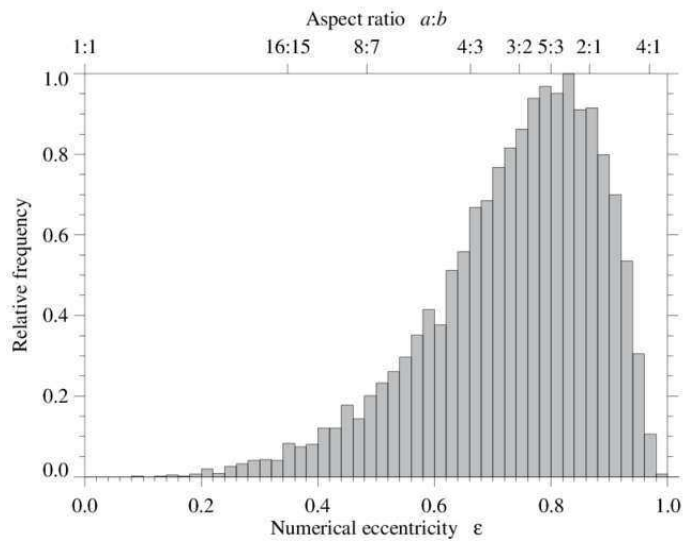}
\caption{Relative frequency distribution of the numerical eccentricity
    $\varepsilon$ of pores derived from ellipse fitting. The aspect ratio
    $a$:$b$
    of the semi-major and -minor axes is provided on the upper scale.}
\label{FIG06}
\end{figure}

\subsection{Area vs. perimeter}

Scatter plots of perimeter length $l_\mathrm{pore}$ vs.\ area
$A_\mathrm{pore}$ are depicted for both data sets in
Fig.~\ref{FIG05}. The ratio $l_\mathrm{pore} / A_\mathrm{pore}$ is minimal for a
perfectly circular pore, i.e., $\alpha = 1$ in \begin{equation} l_\mathrm{pore}
= 2 \alpha \sqrt{\pi A_\mathrm{pore}}. \label{EQN01} \end{equation} All
perimeter lengths have to be above the curve representing circular objects. We
find that the larger the area the farther the points are located from the curve.
To quantify this behavior, we compute the scaling factor $\alpha$ with respect
to this square-root dependence, which describes both the ellipticity and
jaggedness of the pore's perimeter. The results for various area intervals are
presented in Table~\ref{TAB03}. The overall scaling factor is $\bar{\alpha} =
1.43$ and $1.38$ in the interval 0--80~Mm$^2$ for single
and time-averaged G-band images, respectively. Notably, the scaling factor
increases from $\alpha = 1.28$ for the smallest pores (0--10~Mm$^2$) to $\alpha
= 1.85$ for larger pores (30--40~Mm$^2$) observed in single G-band images. The
trend is the same for time-averaged data, where the
scaling factor increases from $\alpha = 1.23$ for the smallest pores
(0--10~Mm$^2$) to $\alpha = 1.54$ for larger pores (30--40~Mm$^2$). To separate
ellipticity from jaggedness, we derive $\alpha_\mathrm{fit}$ from the semi-major
$a$ and -minor $b$ axes of the fitted ellipses (see below) using Ramanjuan's
approximation of the circumference \citep{Villarino2006}. There is a tendency of
small pores to have a more circular shape. However, this effect is small
compared to the fact that larger pores have more corrugated boundaries.
In both data sets only about 20 pores have an area
$A_\mathrm{pore} > 40$~Mm$^2$, so that they are not listed in Table~\ref{TAB03}
and are excluded from Fig.~\ref{FIG05} for clarity. Lastly, the curve for a
scaling factor $\alpha = 2.5$ is a good upper envelope for the scatter plots.

% FIGURE 7
\begin{figure}[t]
\includegraphics[width=\columnwidth]{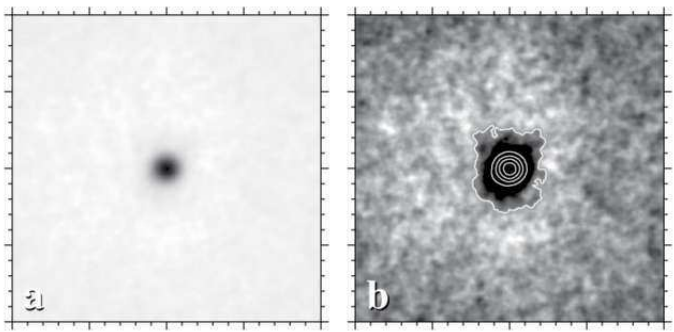}
\caption{The average of 742 single G-band images
    containing isolated pores is displayed between (a) $I_\mathrm{min} =
    0.54\,I_0$ and $I_\mathrm{max} = 1.03\,I_0$ and (b) in the range $(1.00 \pm
    0.02)\,I_0$. The white contour lines correspond to intensities from
    0.6--1.0\,$I_0$ in increments of $0.1\,I_0$. Major tick marks are separated
    by 5~Mm.}
\label{FIG07}
\end{figure}

Figure~\ref{FIG05} also contains the frequency distributions
of perimeter length $l_\mathrm{pore}$ and area $A_\mathrm{pore}$, which are
plotted as gray histogram bars. Their respective parameters are included in each
panel. In addition, log-normal distributions have been fitted to the histogram
\begin{equation}
f\left(x, \mu, \sigma\right) = \frac{1}{x\sigma\sqrt{2\pi}}
    e^{\textstyle - \frac{\left(\ln x - \mu\right)^2}{2\sigma^2}}
    \quad \mathrm{with} \quad x > 0,
\label{EQN02}
\end{equation}
where $\mu$ and $\sigma$ are the mean and standard deviation of the variable
$x$'s (i.e., $l_\mathrm{pore}$ or $A_\mathrm{pore}$) natural logarithm,
respectively. The mean $\mu$ and standard deviation $\sigma$ are given in
Table~\ref{TAB04} along with the moments of the fitted distributions. Mean,
median, and 10$^\mathrm{th}$ percentile are smaller for the fitted distributions
because a significant number of pores larger than 4~Mm$^2$ and with perimeter
lengths in excess of 10~Mm do not follow the log-normal frequency of occurrence.
This relative preponderance of larger pores is also evident in the positive
values of the skewness. A possible explanation for this behavior is the tendency
of the Perona-Malik filter to join nearby pores. In any case, classifying a
chain of pores as either one entity or as separate features is a subjective
decision, as long as no other information (e.g., about the magnetic field)
enters the decision-making process. Nevertheless, the large kurtosis for the
area $A_\mathrm{pore}$ suggests that most pores are small. More than 77\% and
66\% of all pores observed in
single and time-averaged G-band images, respectively, are smaller than
5~Mm$^2$, i.e., they cover a region of just a few granules. In summary, the
overall correspondence between observations and log-normal fits is very good so
that Eqn.~\ref{EQN02} together with Table~\ref{TAB04} can serve as a reference
for further investigations and theoretical/numerical models of solar pores
\citep[e.g.,][]{Leka2001, Cameron2007, Rempel2011c, Hartlep2012}.

\subsection{Eccentricity}

The blob-analysis code \citep{Fanning2011} contains provisions
for ellipse fitting \citep{Markwardt2009} using all pixels within the confines
of the pore's perimeter. Thus, the question can be addressed how circular pores
are. This analysis uses only the larger data set based on
single G-band images. The numerical eccentricity (Fig.~\ref{FIG06}) of an
ellipse is given by
\begin{equation}
\varepsilon = \frac{e}{a} = \frac{\sqrt{a^2 - b^2}}{a}
   \quad\mathrm{and}\quad \varepsilon \in [0,\,1),
\end{equation}
where $e$ is the linear eccentricity and $a$ and $b$ denote the semi-major and
-minor axes, respectively. The aspect ratio of the semi-major and -minor axes
$a$:$b$ is indicated for selected integer ratios, which are sometimes easier to
grasp. The mean $\varepsilon_\mathrm{mean}$ and median
$\varepsilon_\mathrm{med}$ of the numerical eccentricity are 0.74 and 0.76,
respectively, which corresponds to an aspect ratio of about 3:2. Less than 7\%
of all pores have a circular shape with an aspect ratio of less than 8:7, and
slightly more than 26\% of all pores have aspect ratios lower than 4:3. The
tendency of the feature identification algorithm to link some neighboring pores
into chains and sometimes even ring-like structure only partially contributes to
the trend. In general, smaller pores tend to be more circular as already pointed
out when fitting the perimeter length vs.\ area relationship in
Eqn.~\ref{EQN01}.

% FIGURE 8
\begin{figure}[t]
\includegraphics[width=\columnwidth]{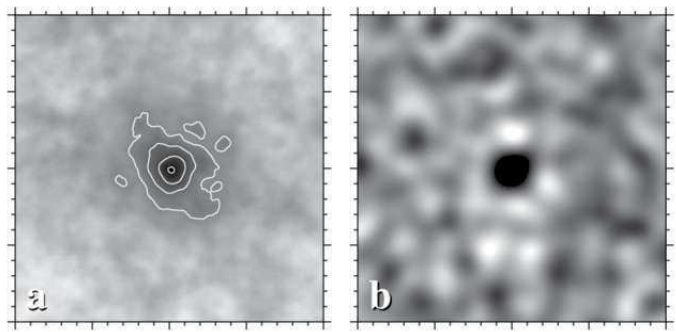}
\caption{Average velocity (a) and divergence (b) of 200
    isolated pores in time-averaged G-band images. The velocity and divergence
    are displayed between 0.15--0.5~km~s$^{-1}$ and $\pm 1 \times
    10^{-4}$~s$^{-1}$, respectively. The white contour lines correspond to
    velocities from 0.15--0.3~km~s$^{-1}$ in increments of 0.05~km~s$^{-1}$.
    Major tick marks are separated by 5~Mm.}
\label{FIG08}
\end{figure}

\subsection{Prototype of an isolated pore}

% Figure 9
\begin{SCfigure*}[1.0][t]
\begin{minipage}{132mm}
\includegraphics[width=66mm]{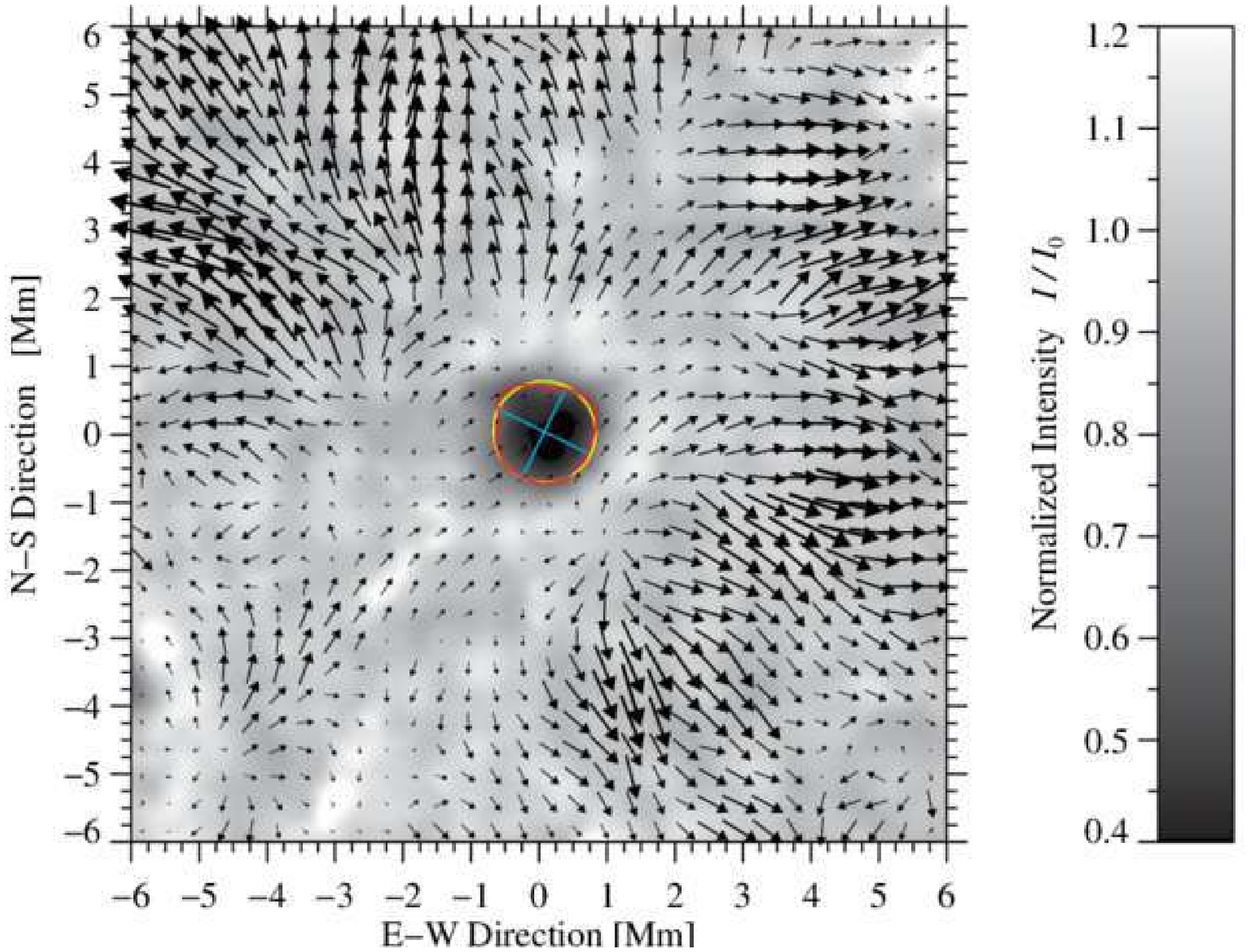}
\includegraphics[width=66mm]{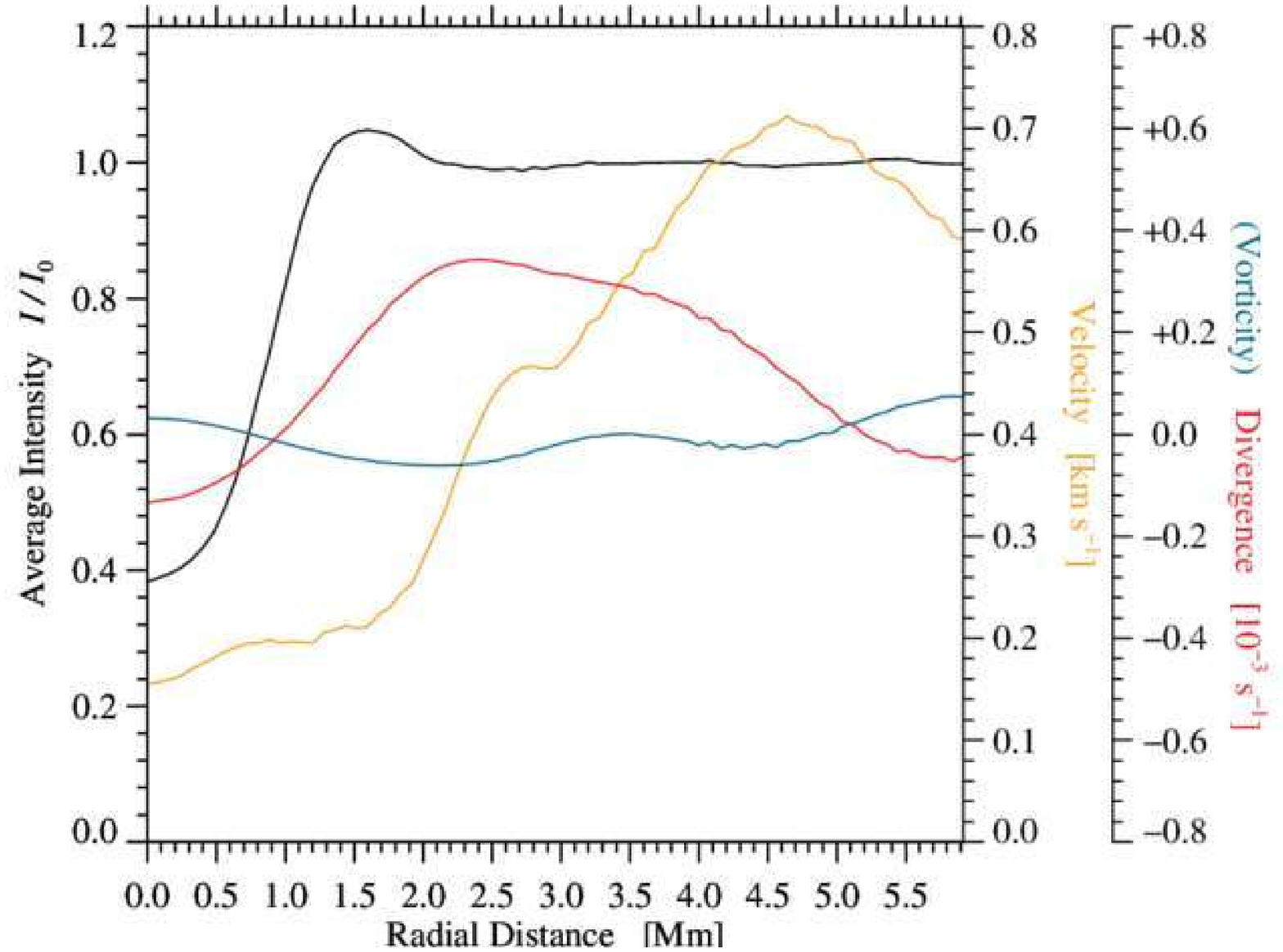}
\smallskip
\includegraphics[width=66mm]{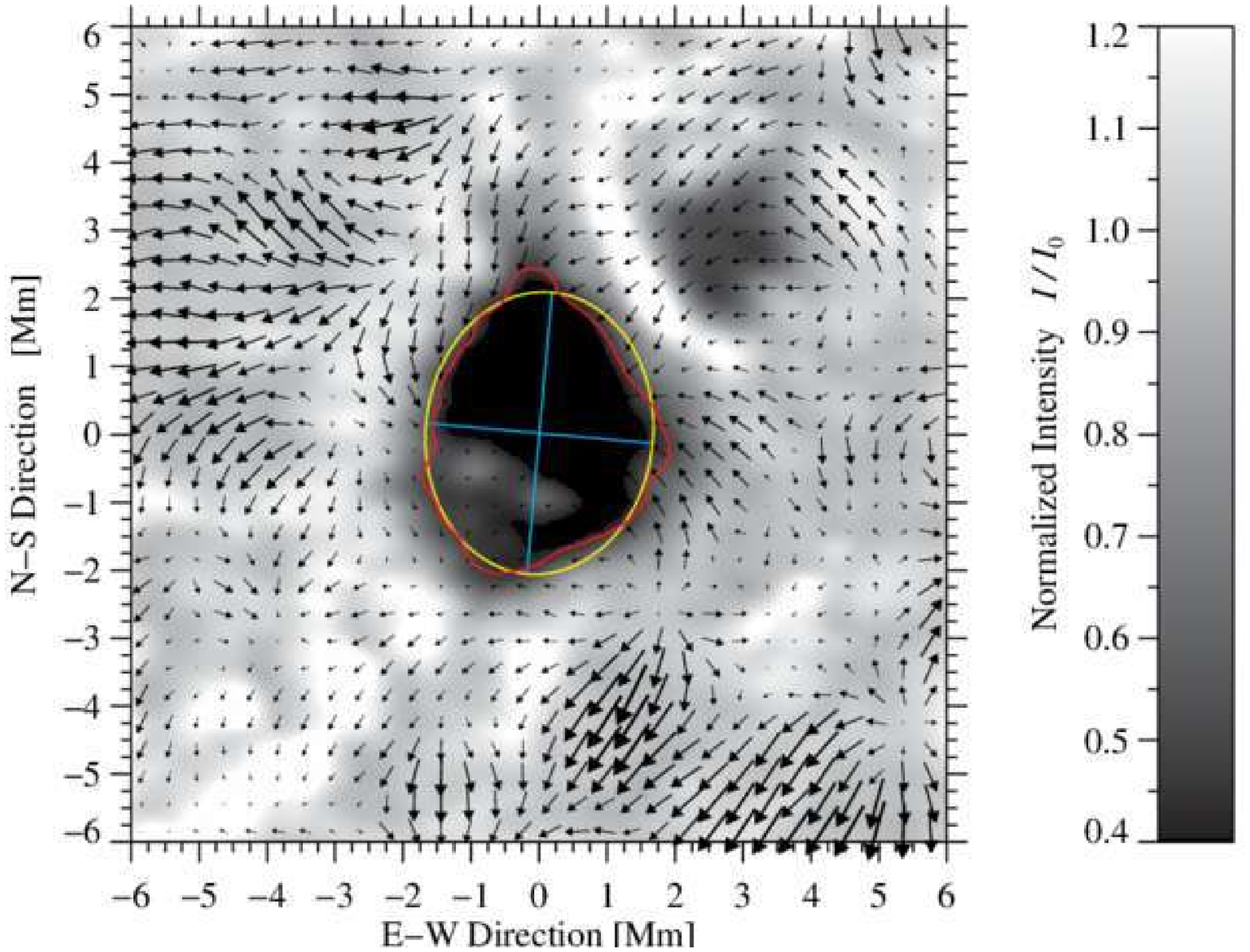}
\includegraphics[width=66mm]{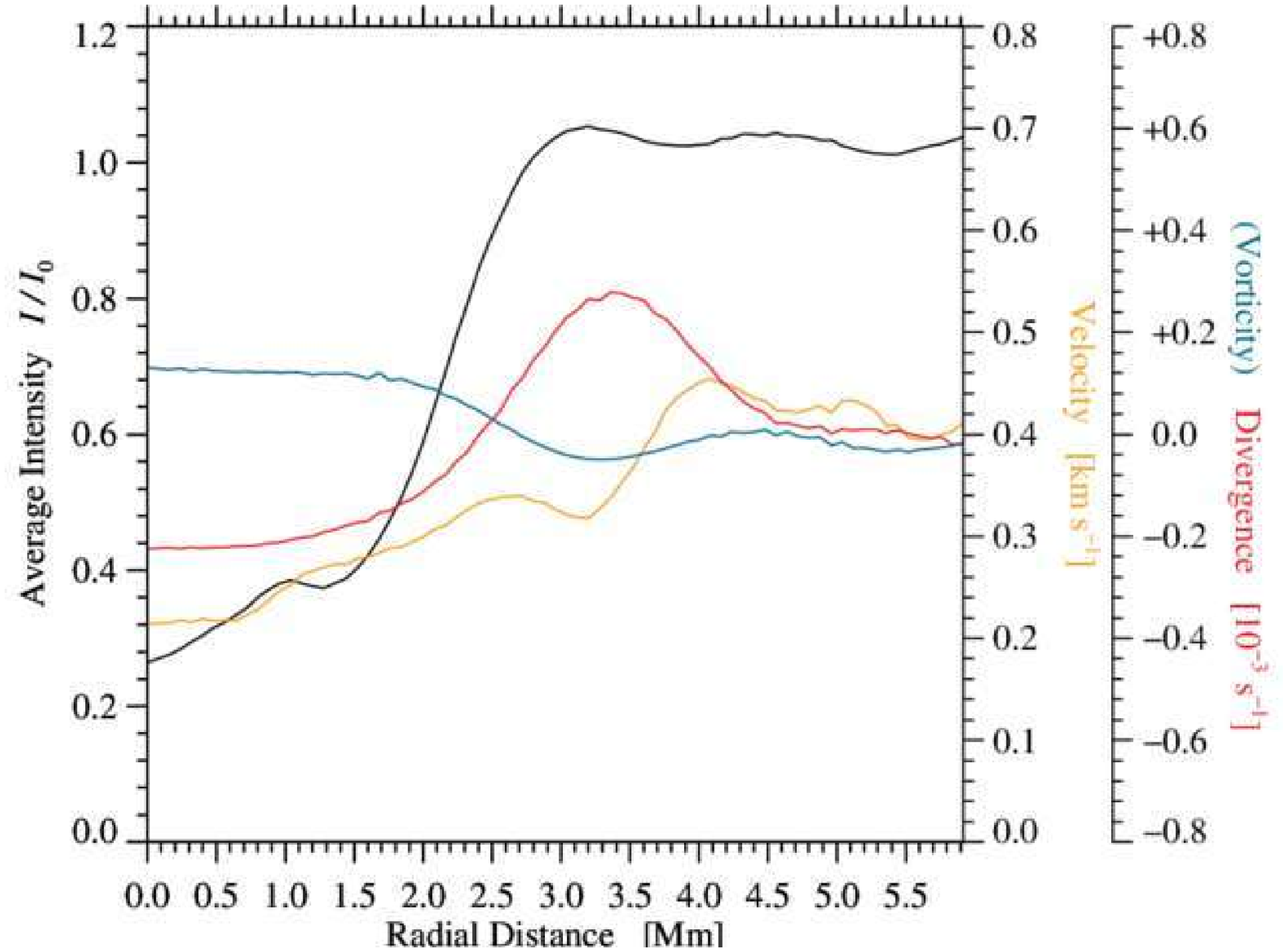}
\end{minipage}
\hspace*{2mm}
\caption{Time-averaged (60-minute) G-band images (\textit{left}) of an isolated
    pore (\textit{top}) observed on 2007 February~3 and a residual pore
    (\textit{bottom}) observed on 2006 December~7. The arrows indicate magnitude
    and direction of the horizontal proper motions. Arrow with the length of the
    grid spacing indicate a speed of 0.5~km~s$^{-1}$. Azimuthally averaged   
    profiles (\textit{right}) are presented for the time-averaged G-band   
    intensity (\textit{black}), the horizontal flow velocities
    (\textit{orange}), the divergence (\textit{red}), and the vorticity
    (\textit{cyan}).}
\label{FIG09}
\end{SCfigure*}

Up to now only scalar parameters have been used to describe
pores. In the next step, radial intensity profiles are computed, which are
representative of a typical pore. Centered on each pore a region-of-interest
(ROI) with a size of 20~Mm $\times$ 20~Mm is extracted from the G-band images.
If the pore is located close to the edge of the images, the ROI is appropriately
trimmed. Strong magnetic features in the vicinity of the pore considerably
reduce the quiet-Sun intensity. Accordingly, these `dark' regions have to be
excluded. Pie-shaped sectors with a width of $1^{\circ}$ are centered on the
pore. If sectors contain any dark feature, they are left aside when computing
the averages. To add a margin of safety, the G-band images are first smoothed by
a Gaussian with a FWHM = 1.68~Mm before an intensity threshold of 0.$85\,I_0$ is
applied to mark contiguous dark regions.

In 742 cases (about 10\% of all pores), no other strong
magnetic feature is detected at all in the ROI of 20~Mm $\times$ 20~Mm. The
arithmetic mean of these ROIs is depicted in Fig.~\ref{FIG07}a scaled between
the minimum ($I_\mathrm{min} = 0.54\,I_0$) and maximum  ($I_\mathrm{max} =
1.03\,I_0$) intensities. The image of the prototypical pore has at first glance
a roundish appearance. Only when displayed in a range of $\pm 2$\% around the
mean quiet-Sun intensity (Fig.~\ref{FIG07}b), more details become visible. The
outermost contour line referring to $I = 1.0\,I_0$ is the only one deviating
from a round or oval shape, because even after averaging it inherits some
structure from pores of very different shapes (see Fig.~\ref{FIG02}). It will be
interesting to see, if this corrugated boundary is related to the magnetic
canopy often observed around pores, which extends well beyond the photometric
boundary of the pores \citep[e.g.,][]{Keppens1996, Leka2001, Balthasar2008}. The
most striking feature in the average image is the granular pattern of G-band
brightenings, which survives the averaging process. These bright features are
clearly visible up to about 5~Mm from the pore's center, and beyond they still
leave an imprint as a faint halo. In conclusion, G-band bright points in
proximity to the pore's boundary are a distinct part of the magnetic flux
system.

Using the same criteria as above, we find 200 isolated
pores in time-averaged G-band images, which is again about 10\% of all pores.
The arithmetic mean of the G-band intensity also reveals the signatures of
bright points in the surroundings of pores. Average horizontal flow speed and
divergence are displayed in Fig.~\ref{FIG08}. We find low velocity and negative
divergence inside pores. The velocity increases monotonously from the center of
the prototypical pore to the surrounding granulation. The most conspicuous
feature in the divergence map is the ring of positive divergence around the
pore. Even after one-hour averaging and computing the arithmetic mean for 200
isolated pores, this ring still survives and asserts the dominance of divergence
centers \citep{Roudier2002} and exploding granules
\citep{VargasDominguez2010}.

\subsection{Flow fields in and around isolated and residual
    pores\label{SECT4.5}}
    
In Fig.~\ref{FIG09}, two pores out of 2863 pores in the
LCT database are presented as an example to explain the data analysis and to
introduce the parameters obtained for further study. We compare two pores with
different histories and backgrounds: an isolated pore in the vicinity of a
sunspot and a residual pore, i.e., the final stage of a decaying sunspot. The
isolated pore is located in a supergranular cell within active region NOAA~10940
observed on 2007 February~3 in the neighborhood of a fully developed sunspot,
whereas the residual pore is the end product of a satellite sunspot
\citep{Verma2012b} in the vicinity of the highly active and flare-prolific
region NOAA~10930 observed on 2006 December~7. We use the time-averaged G-band
image as background and superimpose the averaged flow vectors. Both pores are
surrounded by a bright intensity ring. However, in case of the residual pore,
the bright ring is not regular. An annular structure of positive divergence (not
shown) envelops the bright ring like an onion peel. The positive
divergence structure contains localized divergence centers, which are related to
exploding granules \citep[see][]{Roudier2002}. Pores exhibit inflows in their
interior and outflows at their periphery; both are not necessarily symmetric.
Around the isolated pore outflows are stronger than inflows.

% Table 5
\begin{table}
\begin{center}
\caption{Parameters describing the morphology of pores and the associated flow
field.}\label{TAB05}
\small
\begin{tabular}{lcc}
\hline\hline
Parameters & Isolated & Residual\rule[-3mm]{0mm}{8mm}\\
\hline
$A_\mathrm{pore}$ [Mm$^{2}$]      & \phn3.11    & 16.90 \rule{0mm}{4mm}\\
$l_\mathrm{pore}$ [Mm]            & \phn6.73    & 16.49 \rule{0mm}{4mm}\\
$A_\mathrm{pore}$/$l_\mathrm{pore}$ [Mm]        & \phn1.02    &
\phn1.10\rule{0mm}{4mm}\\
$\varepsilon$                        & \phn0.25    & \phn0.58  \rule{0mm}{4mm}\\
% $\theta$ [\arcdeg]  & $-43.6$     & 85.20 \rule{0mm}{4mm}\\
$I_\mathrm{min}$    & \phn0.37    & \phn0.24  \rule{0mm}{4mm}\\
$I_\mathrm{mean}$   & \phn0.60    & \phn0.48  \rule{0mm}{4mm}\\
$v_\mathrm{mean}$ [km~s$^{-1}$]   & \phn0.18      & \phn0.29  \rule{0mm}{4mm}\\
\mbox{$\nabla\ \cdot v$} [$10^{-3}$~s$^{-1}$] & $-0.05$       & $-0.14$
\rule{0mm}{4mm}\\
$\nabla \times v$ [$10^{-3}$~s$^{-1}$] & \phn0.01 & \phn0.10
\rule[-2mm]{0mm}{6mm}\\
\hline
\end{tabular}
% \footnotesize \hspace*{1mm}
% \parbox{50mm}{\vspace*{-1mm}
% \begin{itemize}
% \item[Note:] 
\tablefoot{The area $A_\mathrm{pore}$ covered by the pore, the length
$l_\mathrm{pore}$ of the circumference, the ratio
$\nicefrac{A_\mathrm{pore}}{L_\mathrm{pore}}$, the numerical
eccentricity $\varepsilon$, the minimum $I_{\mathrm{min}}$
and the mean $I_{\mathrm{mean}}$ intensities, the average flow speed
$v_{\mathrm{mean}}$, the mean divergence \mbox{$\nabla\ \cdot v$}, and the mean
vorticity $\nabla \times v$.}
% \end{itemize}}
\end{center}
\end{table}

% Figure 10
\begin{SCfigure*}[1.0][t]
\begin{minipage}{132mm}
\includegraphics[width=66mm]{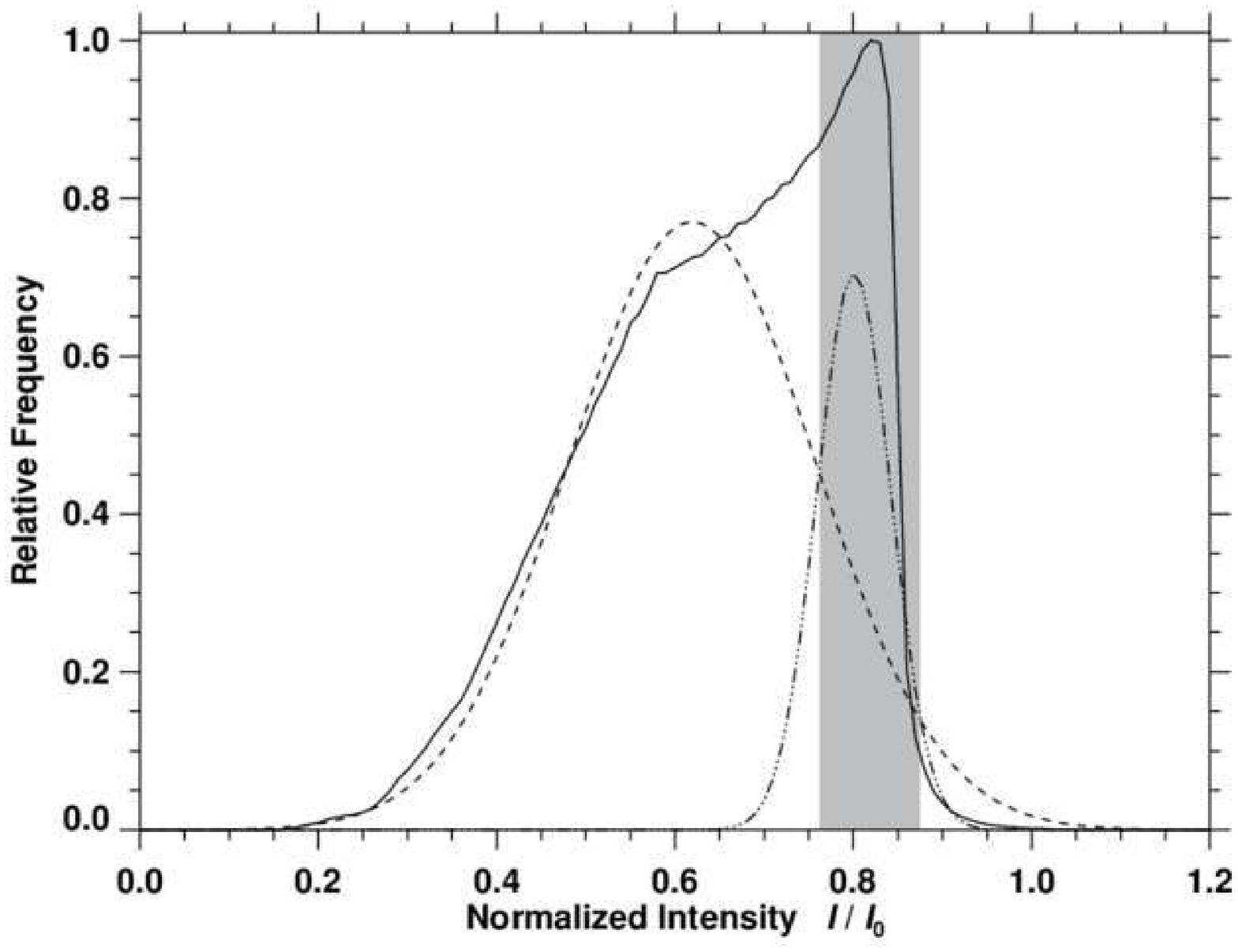}
\includegraphics[width=66mm]{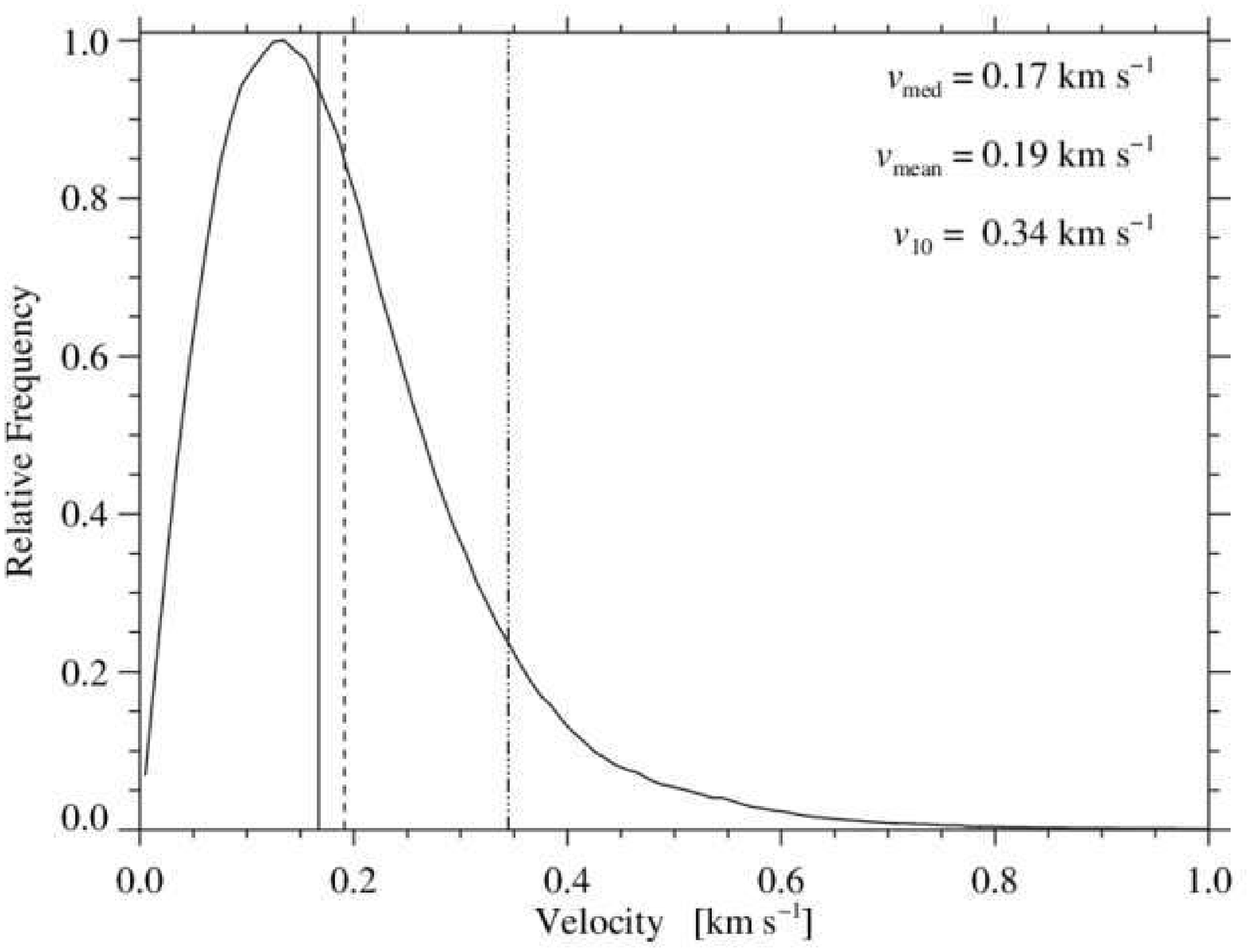}
\smallskip
\includegraphics[width=66mm]{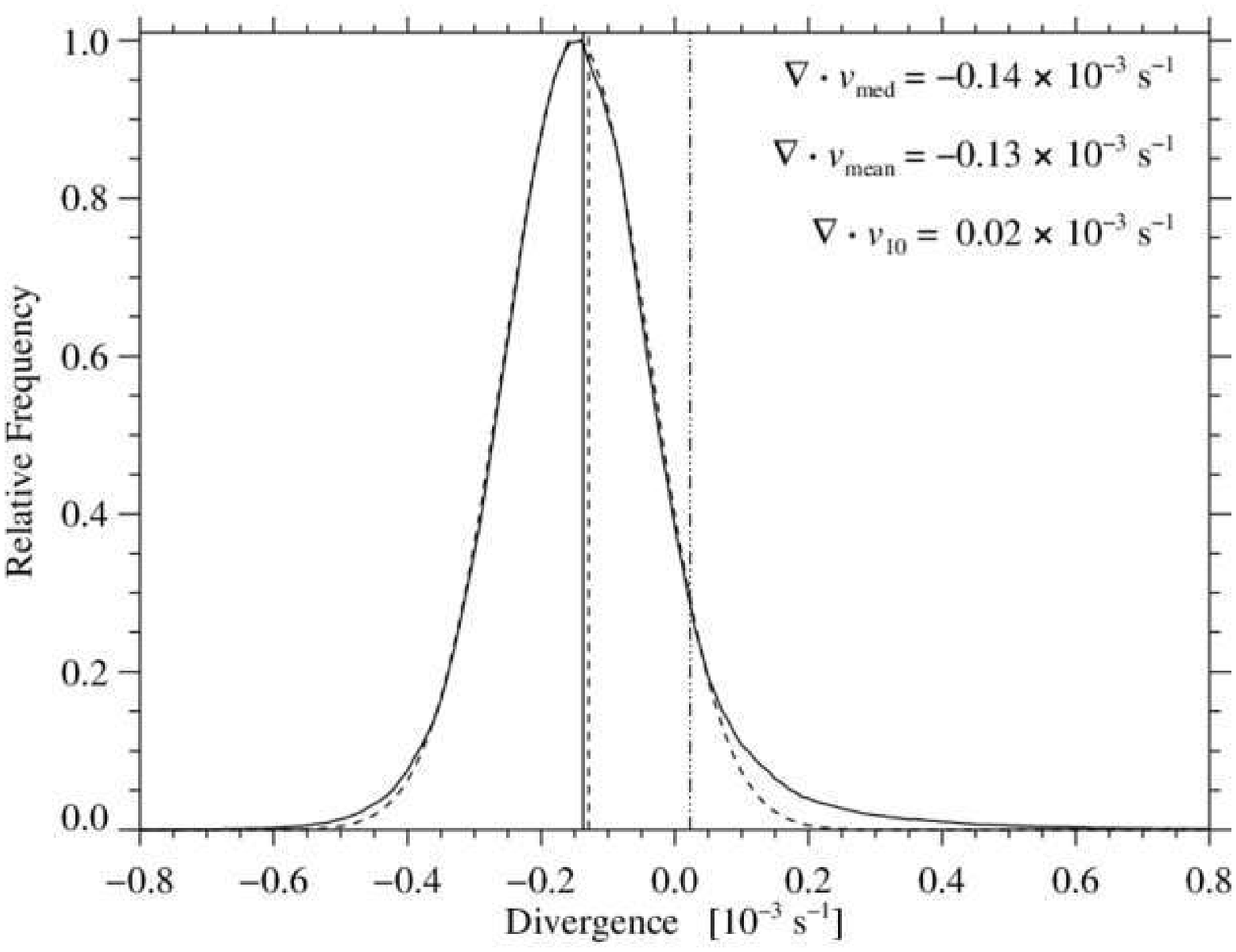}
\includegraphics[width=66mm]{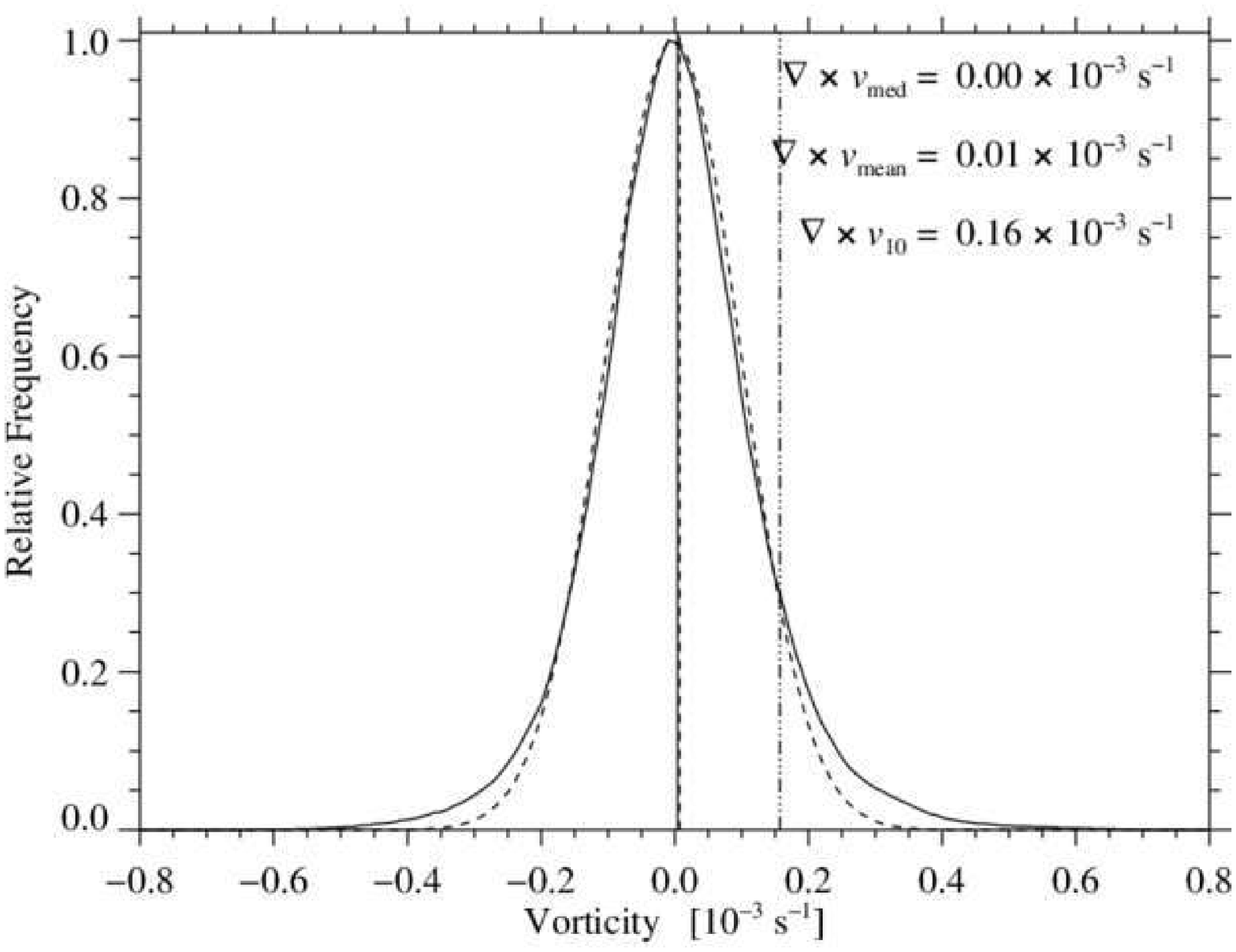}
\end{minipage}
\hspace*{2mm}
\caption{The intensity distribution (\textit{top-left}) of all pores
\mbox{(\textit{solid}\tsp)} can be fitted with two Gaussians, i.e., a dark
component with $\mu_{1} = 0.62\,I_0$ and $\sigma_{1} = 0.14\,I_0$
\mbox{(\textit{dashed}\tsp)} and a bright component with $\mu_{2} = 0.80\,I_0$
and $\sigma_{2} = 0.04\,I_0$ \mbox{(\textit{dash-dotted}\tsp)}.
The gray region outlines the bright component as given
by the optimal thresholds $T_1 = 0.76\,I_0$ and  $T_2 = 0.87\,I_0$. The
frequency distributions are also given for the flow speed (\textit{top-right}),
the divergence (\textit{bottom-left}), and the vorticity (\textit{bottom-right})
within all pores. The three vertical lines mark the position of the median
(\textit{solid}), mean (\textit{long-dashed}), and $10^\mathrm{th}$ percentile
(\textit{dash-dotted}) of the the speed $v$, the divergence \mbox{$\nabla\ \cdot
v$}, and the vorticity $\nabla \times v$. The divergence and vorticity
distributions are fitted with a Gaussian (\textit{dashed}) having a mean value
of $\mu = -0.14\times 10^{-3}$~s$^{-1}$ and $-0.01\times 10^{-3}$~s$^{-1}$,
respectively.}
\label{FIG10}
\end{SCfigure*}

The parameters describing the morphology for both pores are compiled in
Table~\ref{TAB05}. As expected from a visual inspection, the isolated pore has a
smaller eccentricity than the residual pore, i.e., it is more circular. The
minimum intensity $I$ is lower in the residual pore. The mean
divergence within both pores is negative, indicating converging flows as seen in
the superimposed flow vectors. Apart from computing parameters inside pores, we
compute radial averages for intensity, velocity, divergence, and vorticity over
a radial distance of 6~Mm for all pores. Radial averages are only computed, if
pie-shaped segments, which are free of any other dark features, cover at least
$180^{\circ}$ in azimuth. In the right panel of Fig.~\ref{FIG09}, the radial
averages for both pores have been compiled. For both pores, the intensity is low
inside, then monotonically increases, and reaches quiet-Sun levels at the pore's
photometric radius. Just outside the boundary, the intensity reaches a maximum,
which corresponds to the bright ring observed around pores in time-averaged
G-band images. The divergence changes sign at the boundary, and the location of
its maximum is well outside the pores. Interestingly, the divergence maximum
extends even beyond the bright circular ring surrounding pores. The high flow
speed outside the isolated pore is also evident in the radial averages with the
speed reaching values of up to 0.7~km~s$^{-1}$. These two examples befittingly
describe the diversity of pores contained in both data sets.

\subsection{Frequency Distributions}

The relative frequency distribution of the G-band intensity for all pores
detected in time-averaged G-band images is given in
Fig.~\ref{FIG10}. We compute the mean $I_\mathrm{mean} = 0.65\,I_0$, median
$I_\mathrm{med} = 0.67\,I_0$, and $10^\mathrm{th}$ percentile of the frequency
distribution $I_{10} = 0.83\,I_0$. The shape of the
distribution is characterized by a modal value at higher intensities accompanied
by an extended shoulder on the low-intensity side. Interestingly, the frequency
distribution increases almost linearly from 0.58 to $0.80\,I_0$. Even though the
frequency distribution is not bimodal, we fit two Gaussians to it representing
a dark and a bright component. The double-Gaussian fit is
justified because dark pores often show bright intrusions, which are most
commonly the counterparts of umbral dots, but sometimes features similar to
sunspot light-bridges are encountered. The respective frequency distribution is
given by
\begin{equation}
p(I/I_0) =
    \frac{P_1}{\sqrt{2\pi}\sigma_1}\,
    e^{\textstyle -\frac{(I/I_0-\mu_1)^2}{2\sigma_1^2}} +
    \frac{P_2}{\sqrt{2\pi}\sigma_1}\,
    e^{\textstyle -\frac{(I/I_0-\mu_2)^2}{2\sigma_2^2}},
\label{EQN03}
\end{equation}
where $P_{1,2}$ are the probabilities of occurrence with $P_1 + P_2 = 1$. Bright
features cover about 21.6\% of the pore's area, i.e., $P_2 = 0.216$. Means
$\mu_{1,2}$ and standard deviations $\sigma_{1,2}$ of the two Gaussians are
given in the caption of Fig.~\ref{FIG10}. Optimal thresholds $T_{1,2}$
\citep[e.g.,][]{Gonzalez2002} separate the bright and dark components
according to the quadratic equation
\begin{eqnarray}
\lefteqn{AT^2 + BT + C = 0\phn \mathrm{with}}\nonumber \\
 & & A = \sigma_1^2 - \sigma_2^2,\phn
B = 2 \big(\mu_1\sigma_2^2 - \mu_2\sigma_1^2\big),\phn \mathrm{and} \\
 & & C = \sigma_1^2\mu_2^2 - \sigma_2^2\mu_1^2 + 2\sigma_1^2\sigma_2
        \ln\big( \sigma_2 P_1 / \sigma_1 P_2\big).\nonumber
\label{EQN04}
\end{eqnarray}

In this case, two thresholds $T_1 = 0.76\,I_0$ and  $T_2 =
0.87\,I_0$ are required to separate the dark and bright components because of
their strong overlap (see gray background in Fig.~\ref{FIG10}). Combining the
two Gaussians results in a bimodal frequency distribution, which is not
observed. However, the intensities have been obtained over a wide range of
heliocentric angles $\theta$. Both components thus have a distinct CLV in G-band
intensity, which affects their means $\mu_{1,2}(\mu=\cos\theta)$ and standard
deviations $\sigma_{1,2}(\mu=\cos\theta)$. We have carried out numerical
experiments using linear expressions for this functional dependence and
successfully recover the general shape of the distribution in
Fig.~\ref{FIG10}. Likewise the disk coverage of all pores is still not
sufficient to accurately fit the distinctive frequency distribution at hand.
Furthermore, morphological properties and the evolution of pores impact
the distribution.

The frequency distribution based on single G-band images is
virtually identical to the one shown in Fig.~\ref{FIG10}. It includes the G-band
intensity of about five million pixels belonging to 7643 pores with an
equivalent area of more than \mbox{32\;\!000}~Mm$^2$.  The mean, median, mode,
and $10^\mathrm{th}$ percentile of this frequency distribution are
$I_\mathrm{mean} = 0.64\,I_0$, $I_\mathrm{med} = 0.65\,I_0$, $I_\mathrm{mode} =
0.77\,I_0$, and $I_{10} = 0.82\,I_0$, respectively. Again, two Gaussians
represent dark ($\mu_1 = 0.60\,I_0$ and $\sigma_1 = 0.15\,I_0$) and bright
($\mu_2 = 0.78\,I_0$ and $\sigma_2 = 0.055\,I_0$) components. The filling factor
of the bright component is about 20\% ($P_2 = 0.19$). The two thresholds of $T_1
= 0.76\,I_0$ and  $T_2 = 0.86\,I_0$ are almost the same as for pores in
time-averaged G-band images.

The frequency of occurrence is shown in Fig.~\ref{FIG10} for G-band intensity,
flow speed, divergence, and vorticity. The distributions include all pixels
belonging to pores in time-averaged G-band images. The speed distribution is
broad and has a high-velocity tail. The mean speed is low $v_\mathrm{mean} =
0.18$~km~s$^{-1}$ inside pores. However, because LCT tracks intensity contrasts,
the lack of contrast-rich structures in the dark core potentially leads to lower
speeds. The frequency distributions of divergence and vorticity have almost the
same shape. We fit the divergence distribution with a Gaussian
\begin{eqnarray}
f\left(x, \mu, \sigma\right) = \frac{1}{\sigma\sqrt{2\pi}}
    e^{\textstyle - \frac{\left(x - \mu\right)^2}{2\sigma^2}}
    \quad,
\label{EQN06}
\end{eqnarray}
with a standard deviation of $\sigma = 0.11\times 10^{-3}$~s$^{-1}$ and a mean
of $\mu = -0.14\times 10^{-3}$~s$^{-1}$. The negative mean value of the
divergence inside the pore indicates inflows, which we also see in the two pores
presented as examples in Fig.~\ref{FIG09}. A normal distribution is likewise a
valid representation of the vorticity with a mean close to zero and
a standard deviation of $\sigma = 0.10\times 10^{-3}$~s$^{-1}$. This indicates
probably that on average neither twisting nor spiraling motion are present
inside pores.

\subsection{Radial averages}

To express the average properties of intensity and flow fields around pores in
more quantitative terms, we compute for each instance azimuthal averages of
intensity, velocity, divergence, and radial component of velocity.
Pores come in a variety of shapes so that computing meaningful
azimuthal averages becomes a challenging task. In the
following, we describe two different approaches to compute azimuthal averages
based on single and time-averaged G-band images. 

\begin{figure}[t]
\includegraphics[width=\columnwidth]{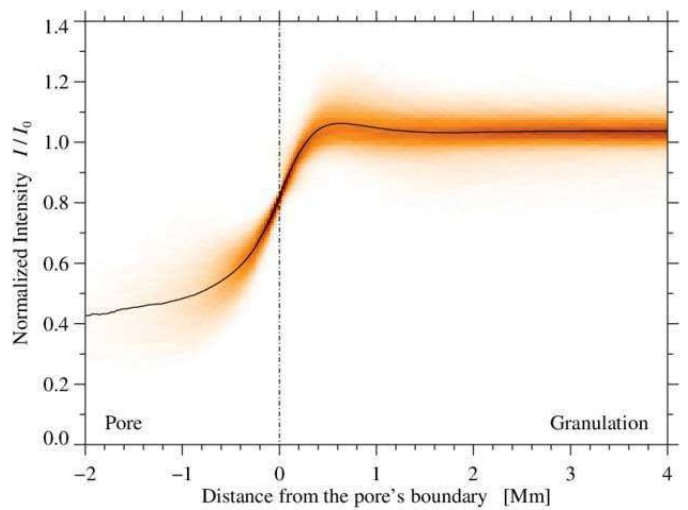}
\caption{Typical intensity profile of a pore  based on single G-band images
    after azimuthal averaging. The
    vertical dash-dotted line indicates the boundary of the pore corresponding
    to the intensity threshold of $0.85\,I_0$. The background image shows the
    two-dimensional frequency distribution of the intensities.}
\label{FIG11}
\end{figure}

Morphological dilation/erosion using disk-shaped binary masks
of increasing/decreasing size is one option to tackle this problem as
demonstrated for single G-band images (see
Fig.~\ref{FIG11}). These operations are applied to binary masks outlining each
pore such that they either expand or shrink the region. Subtracting a region
from the next smaller one results in a ring-like mask, which still preserves the
shape of the pore. The ring-like structure becomes more circular for larger
distances from the pore's origin, which is a desired feature for an azimuthal
average. Taking the mean of the positions and intensities along the circular
ring of pixels yields a radial intensity profile, which is resampled to a
regular grid. This allows us not only to compute a global mean of all intensity
profiles but also to derive the intensity distribution at each radial position.

Distances are measured from the pore's edge in
Fig.~\ref{FIG11}. We find a preponderance of G-band brightenings in proximity
(about 0.64~Mm) to the pore's boundary (vertical dash-dotted line). G-band
bright points are located well within the ring-like structure of divergence
centers \citep[e.g.,][]{Roudier2002} at a distance of about 3.5~Mm from the
pore's boundary. \citet{Sobotka1999} relate recurring G-band bright points in
the vicinity of pores to mesogranular flows \citep{November1988}. The intensity
gradient is $dI / dr \approx 0.7\,I_0$ Mm$^{-1}$ in the linear part of the
radial intensity profile. Using the edge of the pore as point of reference has
the advantage of describing the intensity inside a pore regardless of its size.
The frequency distribution first flares out and then tapers off, because pores
with radii larger than 1~Mm are rare. The tendency that larger pores are darker
\citep[e.g.,][]{Keppens1996} is evident but the interior gradient $dI /
dr\approx 0.06\,I_0$ Mm$^{-1}$ is shallow. The narrowing of the frequency
distribution at the edge of the pore is an artifact of the intensity
thresholding of the Perona-Malik filtered images. That the radial intensity
profile does not reach unity for large distances from the pore is because of the
asymmetric frequency distribution with an extended tail towards higher
intensities and the faint halo of G-band brightenings seen in
Fig.~\ref{FIG07}b.

% FIGURE 11

% Figure 12
\begin{SCfigure*}[1.0][t]
\begin{minipage}{132mm}
\includegraphics[width=66mm]{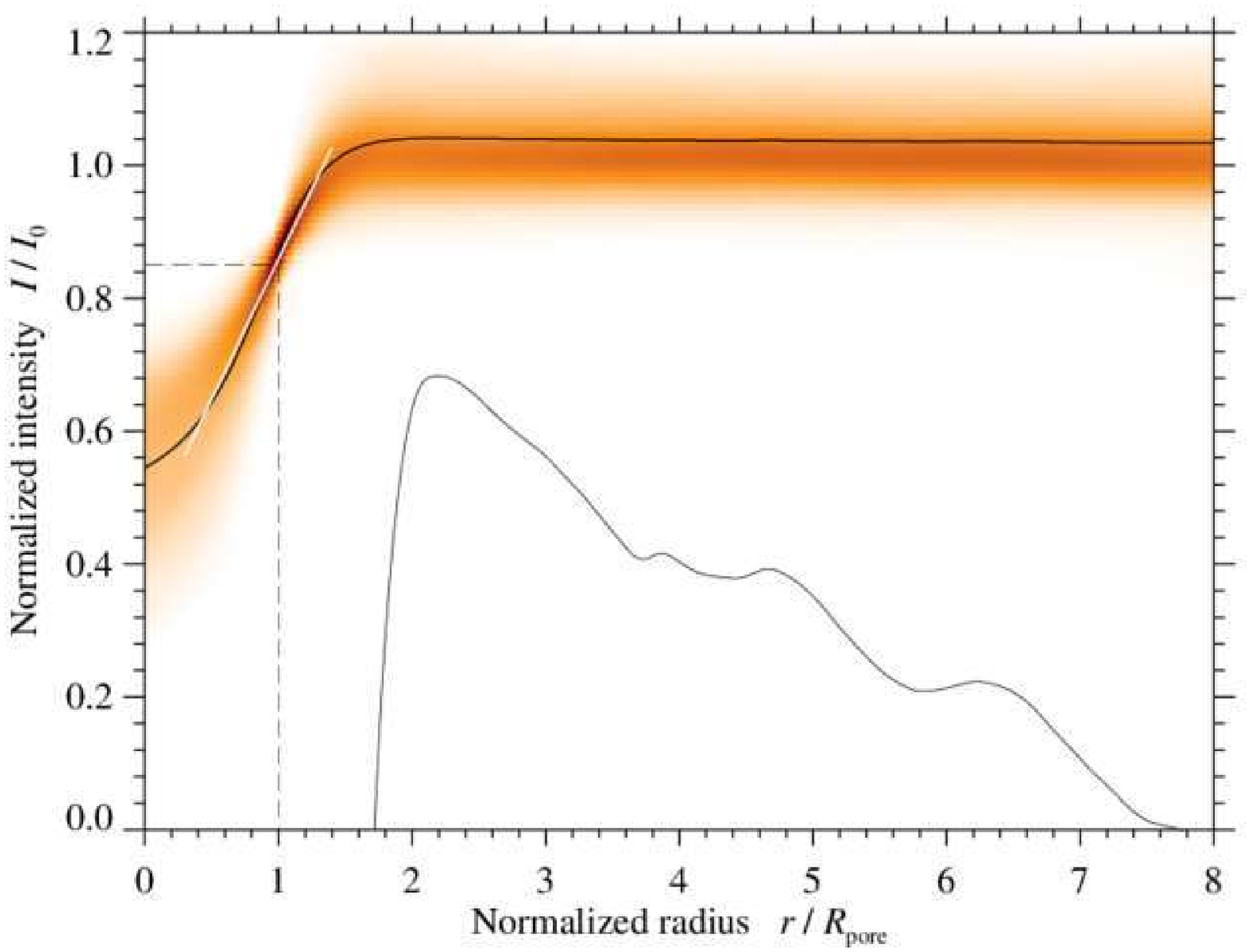}
\includegraphics[width=66mm]{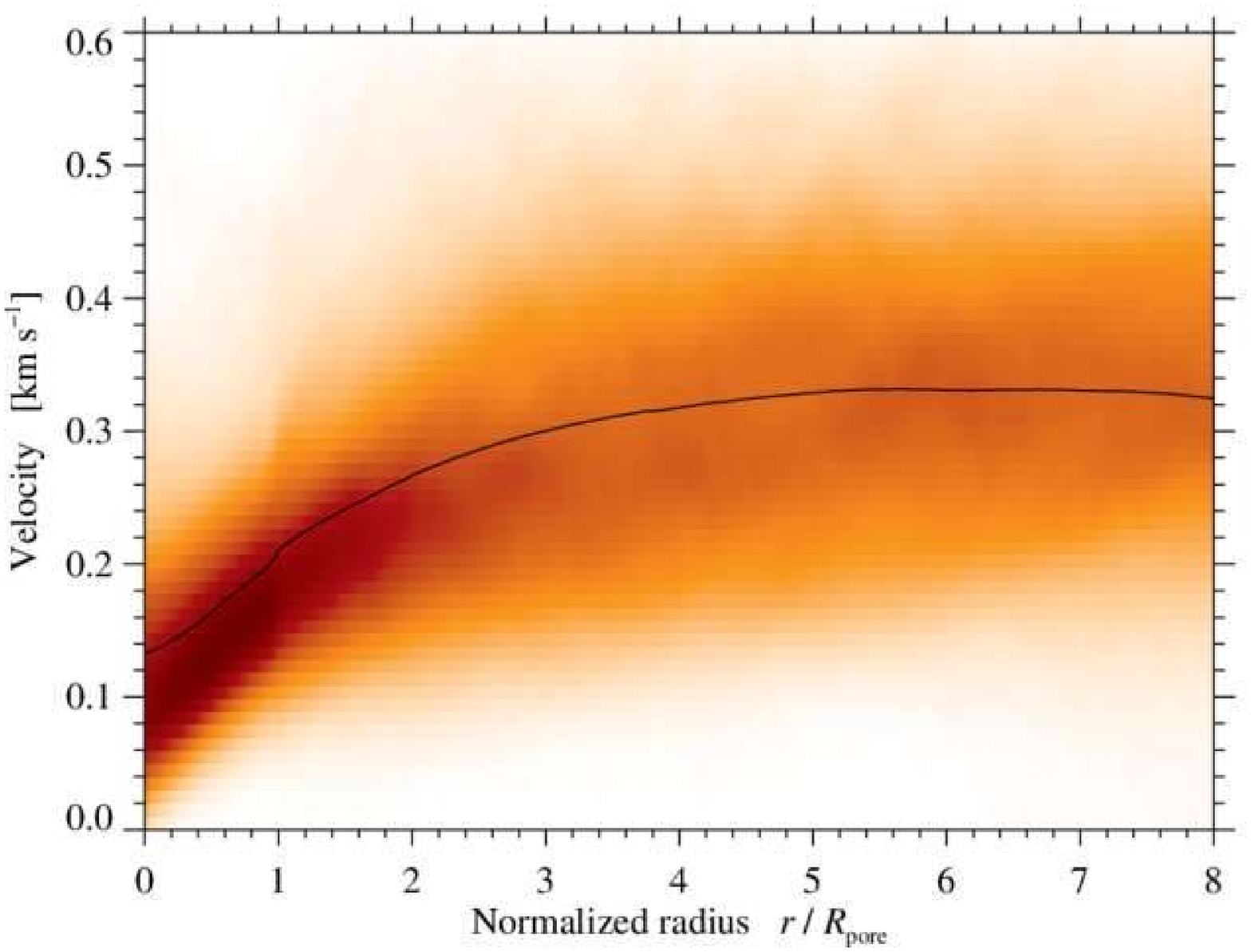}
\smallskip
\includegraphics[width=66mm]{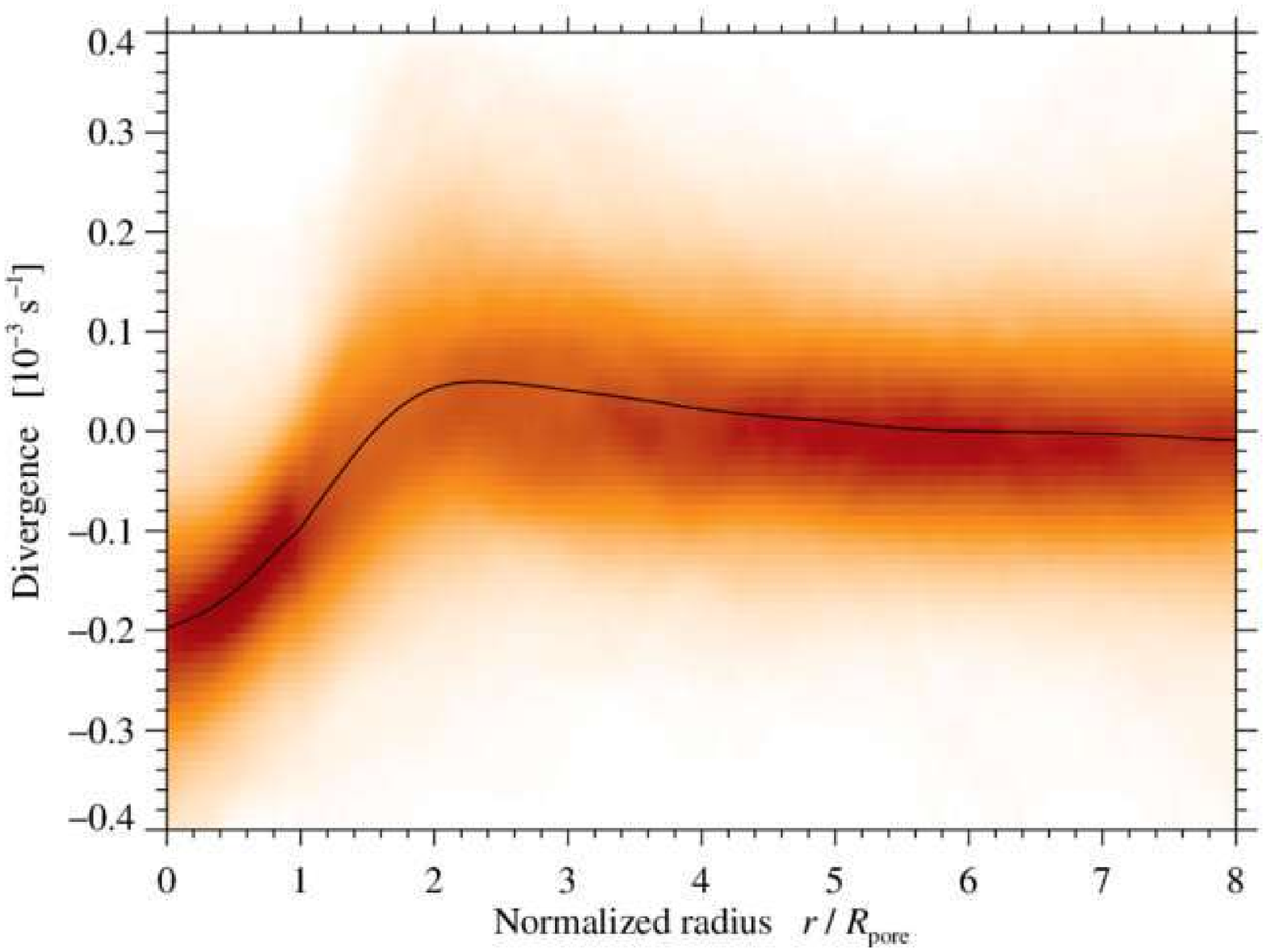}
\includegraphics[width=66mm]{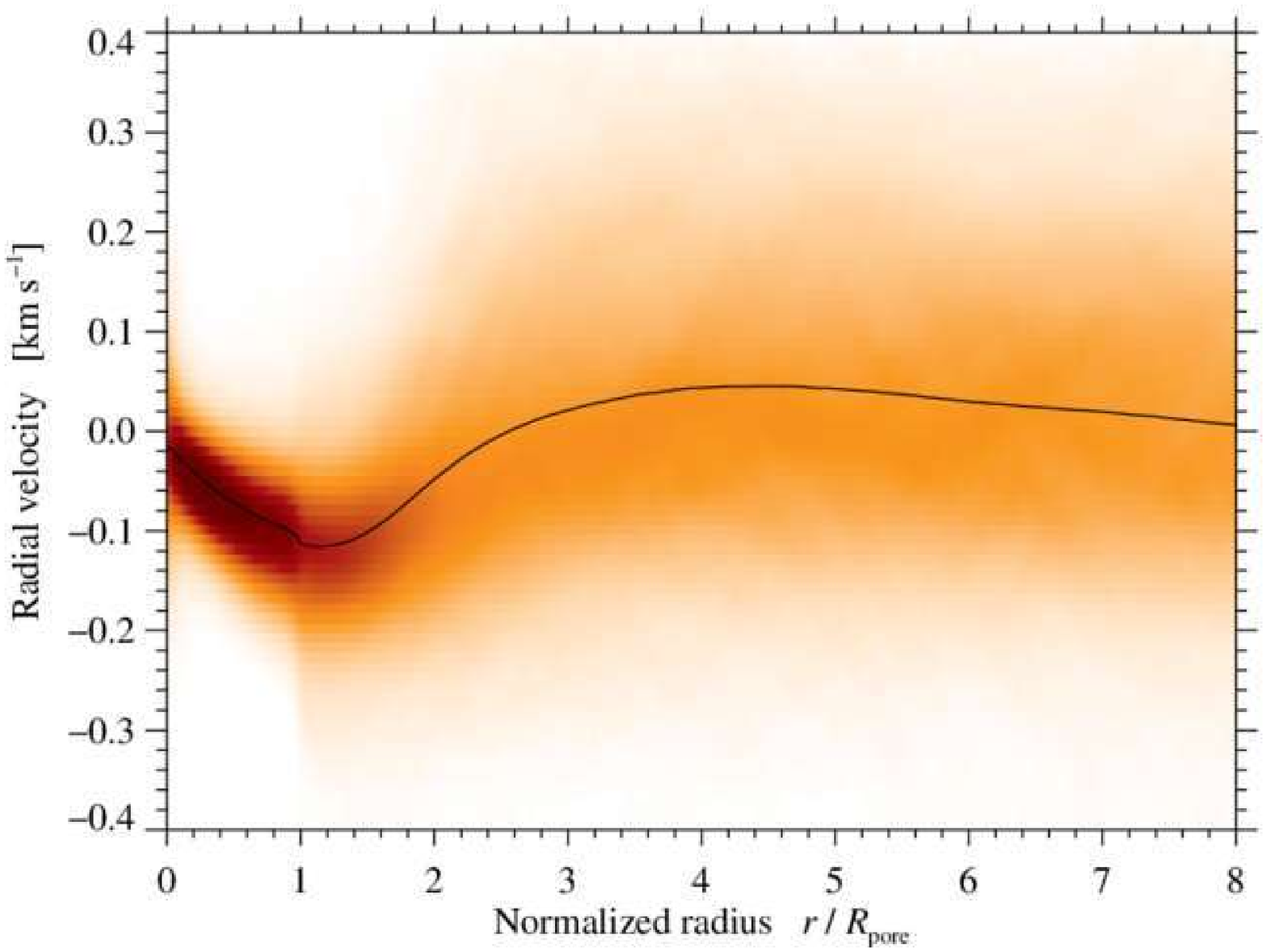}
\end{minipage}
\hspace*{2mm}
\caption{Azimuthally averaged profiles for intensity (\textit{top-left}), flow
    velocity (\textit{top-right}), divergence (\textit{bottom-left}), and radial
    velocity (\textit{bottom-right}) for all pores. The profiles are computed
    for normalized radial distances, where \mbox{$\nicefrac{r}{R_\mathrm{pore}}
    = 0$} and \mbox{$\nicefrac{r}{R_\mathrm{pore}} = 1$} indicate the center and
    the boundary of the pores, respectively. The frequency of occurrence is
    shown as a shaded background,  where darker regions refer to more commonly
    encountered values. In the intensity profile (\textit{top-left}), the
    \textit{solid} curve represents the percentage of increase in average
    intensity, and the \textit{dashed} line reflects our choice of the intensity
    threshold (0.85\,$I_0$) delimiting pores. The \textit{white} line
    corresponds to an intensity gradient of $dI / dr =
    0.42\,I_0\,R_\mathrm{pore}^{-1}$.}
\label{FIG12}
\end{SCfigure*}

An alternate approach has been employed to compute
azimuthal averages for pores contained in time-averaged G-band images. Starting
with more than 2863 pores we deselect pores, which are at FOV's border, as their
radial averages rely on a much smaller number of points. Thus, we are left with
2300 pores. The averages are computed for a FOV of 300 $\times$ 300~pixels
(24~Mm $\times$ 24~Mm) centered on the pore's origin. Pores are not always
circular. Hence, considering their irregular shapes while computing azimuthally
average profiles outside and inside pores, we grow/shrink the region
successively by 100/50 pixels using morphological dilation/erosion. Shrinking of
a region is stopped, once the origin is reached. In addition, we exclude any
$1^{\circ}$-wide sector that contains either dark pores or any other dark
feature. To facilitate the comparison of all pore profiles, we interpolate
individual profiles on a regular grid of normalized radius $R_\mathrm{N} =
\nicefrac{r}{R_\mathrm{pore}}$. These averaged profiles are depicted in
Fig.~\ref{FIG12}. We also compute frequency distributions at each normalized
radial position for every physical parameter. The distributions are included in
the panels as shaded backgrounds, where darker gradations of color represent a
larger frequency of occurrence.

In the averaged intensity profile, the intensity is low inside the prototypical
pore and slowly approaches a level slightly larger than \mbox{$I / I_0 = 1$}.
The spread in intensity values (shaded background) is small and the mode
conforms more closely to \mbox{$I / I_0 = 1$}. The intensity has a gradient of
$dI / dr = 0.42\,I_{0}\,R_\mathrm{N}^{-1}$ in the linear part of the radial
intensity profile as indicated by a white-line in Fig.~\ref{FIG12}. The
intensity peak, which we see in the azimuthal profile of two pores
(Fig.~\ref{FIG09}), is much harder to discern in the averaged intensity profile.
To clearly identify the peak in intensity, we also plot the excess brightness
with respect to the mean intensity at the FOV's periphery. These values were
multiplied by a factor of hundred. Thus, the scale in Fig.~\ref{FIG12} can be
interpreted as the excess brightness in percent. As evident from the plot, there
is an increase of about 0.7\% in intensity around the border of pores (at about
2~$R_\mathrm{N}$) indicating that on average pores are surrounded by many G-band
bright points. Despite using two different approaches, the
overall appearance of the azimuthally averaged intensity profiles is virtually
the same, which confirms that both approaches are effective in deducing the
physical properties of pores. 

The average velocity profile does not show any peculiarities. Inside the pores
the velocities are low, then they monotonically increase until reaching
quiet-Sun values. However, the velocity distributions as a function of
normalized radius are narrower out to about three times the pore's radius. We
attribute this compactness and the lower mean flow speed to the presence of a
magnetic canopy around pores \citep[e.g.,][]{Suetterlin1996}, which inhibits to
some extend the horizontal plasma flows. The mean velocity outside pores
$R_\mathrm{N} > > 1$ is about 0.31~km~s$^{-1}$. The median is 0.32~km~s$^{-1}$
and the 10$^{\mathrm{th}}$ percentile amounts to 0.33~km~s$^{-1}$. To gain more
insight how the flow speed varies around pores, we compute the radial component
of horizontal flow velocity using $v_\mathrm{rad} = v_{x} \cos \theta + v_{y}
\sin \theta$, where $\theta$ is the azimuth angle measured counterclockwise from
the positive $x$-axis around the center of the pores. The radial velocity starts
close to zero ($-0.01$~km~s$^{-1}$) inside the pore, then decreases to reach a
minimum of $-0.12$~km~s$^{-1}$ at around $R_\mathrm{N} = 1.1$, before increasing
again attaining positive values at $R_\mathrm{N} = 2.6$. Finally, it approaches
its maximum (0.05~km~s$^{-1}$) at around $R_\mathrm{N} = 4.6$ before leveling
out to zero far away from the pore's origin. The width of the frequency
distributions is largest near $R_\mathrm{N} = 4$.

The divergence is negative inside pores and switches to positive values at
around $R_\mathrm{N} = 1.6$ before attaining a maximum of 0.05 $\times
10^{-3}$~s$^{-1}$ at $R_\mathrm{N} = 2.3$. Once the maximum is reached, the
divergence slowly converges on zero. The divergence peak at around $R_\mathrm{N}
= 2.3$ is also clearly visible in the width of the frequency distribution. This
divergence peak has already been discussed in Sect.~\ref{SECT4.5} in the context
of the isolated and residual pores. A negative divergence indicates inflows
inside pores and positive values surrounding pores suggest outward motions near
the borders. The zero crossing is located at $R_\mathrm{N} = 4$.

To summarize our findings, we compile all radially averaged profiles in
Fig.~\ref{FIG13} to furnish a complete picture of the photometric and flow
characteristics around a prototypical pore. We have chosen an intensity
threshold of 0.85\,$I_0$ to identify pores, and all the results discussed so far
refer to this choice. Now, the question arises why not use an intensity
threshold of 1.0\,$I_0$ for the boundary of a pore. The contrast between bright
granules and dark intergranular lanes in G-band images renders such an approach
futile leading to ill-defined boundaries for individual pores. Only after
averaging over many pores and taking an azimuthal average, a `photometric
radius' of about 1.4 $R_\mathrm{N}$ can be determined when approaching the mean
quiet-Sun intensity. This is about 40\% larger than our original definition of a
pore's boundary. Hence, all linear dimensions have to be scaled by 40\% when
referring to the photometric radius.

Moving outwards from pore's border, we find an intensity peak around
$R_\mathrm{N} = 2.1$, which is the average location of the encircling G-bands
bright points. Inspecting the horizontal flows, as expected, the radial
component of the horizontal velocity $v_\mathrm{rad}$ is zero at the center of
the pore. The strongest inflows, i.e., the minimum of $v_\mathrm{rad}$ is
observed at the location $R_\mathrm{N} = 1.1$, which is virtually identical to
our choice of the pore's boundary (intensity threshold of 0.85 $I_0$). The
radial velocity changes sign at $R_\mathrm{N} = 2.6$ slightly outside the
intensity maximum (ring of G-band brightenings), which indicates that horizontal
inflows start well outside the pore's border. The radial flow field of pores
reaches significantly outward with a maximum of $v_\mathrm{rad}$ at
$R_\mathrm{N} = 4.6$ before approaching zero at a spatial scale corresponding to
the supergranular cell size. To identify sources and sinks of the flow fields in
the vicinity of pores, we compute the divergence \mbox{$\nabla \cdot v$}. The
divergence is zero at $R_\mathrm{N} = 1.6$, which corresponds to the photometric
radius $R_\mathrm{N}=1.4$. Hence, we can define the photometric radius also as
the location, where the divergence \mbox{$\nabla \cdot v = 0$}. The maximum of
the divergence is located between the maxima of intensity and radial velocity,
which represent two rings encircling pores like an onion skin
\citep[e.g.,][]{Sobotka1999, Roudier2002}. However, the in- and outflows in
pores are easier to grasp employing the radial component of the horizontal
velocity. The divergence has contributions from both radial and tangential
velocity components, whereas $v_\mathrm{rad}$ is just the velocity vector
projected onto the unit vector in the radial direction.

%###############################################################################
%#
%#    CONCLUSIONS
%#
%###############################################################################

\section{Conclusions}

Pores represent two stages in the evolution of sunspots, either as a dark umbral
core forming a penumbra when growing to become a sunspot or as a decaying
sunspot loosing its penumbra. Many pores, however, never reach this transient
period. Regardless of the pores' evolutionary phase, their general statistical
properties still improve our understanding of how sunspots form and decay.

% Figure 13
\begin{figure}[t]
\includegraphics[width=\columnwidth]{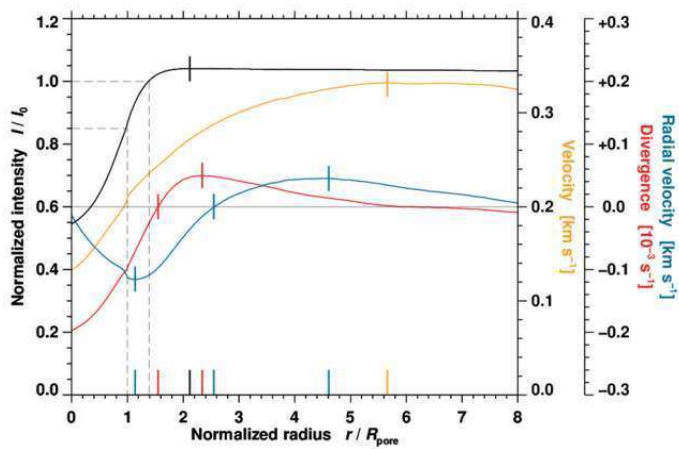}
\caption{Azimuthally averaged profiles of intensity (\textit{black}), flow
    velocity (\textit{orange}), divergence (\textit{red}), and radial velocity
    (\textit{green}) for all pores (same as Fig.~\ref{FIG11} using a similar
    layout as in Fig.~\ref{FIG08}). Two \textit{dashed} line represent our
    choice of intensity threshold (0.85\,$I_0$) and the photometric radius of
    pores (1.0\,$I_0$). The \textit{solid} line in the center is provided as a
    visual guide for the zero-crossing of divergence and radial velocity. The
    short colored vertical lines mark the extreme values and zero-crossings.
    These markers are replicated on the abscissa to assist in finding the 
   corresponding normalized radii.} 
\label{FIG13}
\end{figure}

All previous statistical studies of pores have been
campaign-driven and have not been designed as systematic surveys. We present the
first comprehensive statistical study of pores concerning their morphology,
photometric properties, and associated horizontal flow fields. This survey makes
use of the extensive database of the Japanese Hinode mission. We have tried to
minimize potential biases, but we have to admit that the study is certainly not
bias-free. The target selection based on `Hinode Observing Programs' favors
flare-prolific active regions, i.e., sunspot groups with complex magnetic field
configurations and (multiple) tangled magnetic neutral lines. On the other hand,
a large number of Hinode quiet-Sun observations is void of any pores indicating
that isolated pores are surprisingly rare. They only amount to about 10\% of the
pores in the sample.

The newly created LCT database \citep{Verma2011} facilitates statistical studies
of flows around solar features in a straightforward manner. In total, 7643/2863
pores (single/time-averaged G-band images)
contribute to the current analysis. We have computed perimeters and areas
describing the morphology of pores, and we provide the radial averages for
intensity, velocity, divergence, and radial component of
the horizontal velocity. Two pores have been presented as an example to
illustrate the data analysis strategy and the types of parameters available for
further examination. Despite its broad scope, the statistical investigation of
pores has not yet reached its culmination. Other physical parameters, most
importantly the magnetic field and Doppler velocities, and the temporal
evolution of pores have to be included. In principle, this can be accomplished
by using synoptic SDO data with excellent coverage but at the moment, such work
has to be deferred until a future time.

In summary, we recite the most important findings based on morphology,
photometric properties, and horizontal proper motions:

(1) A pronounced north-south asymmetry is observed in the
distribution of pores on the solar disk (Fig.~\ref{FIG03}). During the extended
minimum of solar cycle 23, pores predominantly appear at low latitudes in the
southern hemisphere. However, this trend changes with the rise of the new solar
cycle 24, when pores are more commonly encountered at high latitudes in the
northern hemisphere (Fig.~\ref{FIG04}).

(2) In general pores are not circular. Ellipse fitting
(Fig.~\ref{FIG06}) yields a typical aspect ratio of 3:2 between semi-major and
minor axes. Only 7\% of all pores identified in single G-band images have a
close-to-circular shape with an aspect ratio smaller than 8:7.

(3) Most pores are small. About 75\%/66\%
(single/time-averaged G-band images) of all pores are smaller than 5~Mm$^2$
(Fig.~\ref{FIG05}), which is comparable to an area covered by just a few
granules. Interestingly, smaller pores tend to be more circular and have less
corrugated boundaries. This is partially explained by the tendency of larger
pores to align in chains and sometimes even in ring-like structures. The feature
identification algorithm does not always separate elongated features into their
constituent parts.

(4) Averaging all 742 isolated pores detected in single
G-band images results in, at least at first glance, a roundish prototypical
pore. Only after narrowing the contrast range to $\pm$2\% around quiet-Sun
intensity, this representative pore exhibits a corrugated boundary surrounded by
a bright, granular intensity distribution brought about by averaging a multitude
of G-band brightenings. Thus, G-band bright points encircling pores constitute a
distinct part of the magnetic flux system. Furthermore, contours of the pore
close to quiet-Sun intensity levels are oval with the longer axis in the
north-south direction. It is tempting to relate the distinct east-west asymmetry
of the bright halo depicted in Fig.~\ref{FIG07} to that of the moat around
sunspots \citep{Sobotka2007}. However, this lopsidedness is neither found in
intensity nor in the velocity and divergence maps (Fig.~\ref{FIG08}) of the
representative pore based on time-averaged G-band images.

(5) The intensity distribution of the pore's interior has a distinct shape
(top-left panel of Fig.~\ref{FIG10}), which can be reproduced by two Gaussians
representing dark and bright components. A double peak is absent because both
mean and width of the Gaussians depend on the heliocentric angle $\theta$. This
dependency is expected to be stronger for bright features (umbral dots and
light-bridges), if the CLV of G-band bright-points is taken as a guide.

(6) The averaged, radial intensity profiles of 7643/2863 pores
show a distinct peak (Fig.~\ref{FIG11} and top-left panel of Fig.~\ref{FIG12}),
which comes out more clearly in radial profiles of individual pores. This peak
indicates the presence of bright features encircling pores. Because the origin
of G-band bright points is related to the physics of small-scale flux tubes
\citep[e.g.,][]{Steiner2001, Schuessler2003}, it would be hasty to conclude that
the bright ring is due to the convective energy blocked by the strong magnetic
fields inside the pore.

(7) The frequency distributions of divergence and vorticity within pores
are easily fitted with Gaussian frequency distributions (bottom panels in
Fig.~\ref{FIG10}). The divergence is predominantly negative within pores
strongly suggesting inflows. The mean value of the vorticity is about zero in
the pore's interior pointing to the absence of twisted or spiraling motions.

(8) The radial component of horizontal velocities is negative inside pores and
approaches positive values outside pores (Fig.~\ref{FIG13}). The negative values
inside imply inflows inside while outflows start about two radii from the pore's
center.

(9) Inflows and outflows seen in the radial velocity profile can also be traced
in the divergence profile. The maximum of the radial divergence radial profile
matches the zero-crossing of the radial velocity profile (Fig.~\ref{FIG13}). The
ring of positive divergence encircling pores corresponds
to the location of exploding granules \citep{Keil1999, Roudier2002,
VargasDominguez2010}.

(10) The photometric radius of the prototypical pore is the location where the
pore's intensity profile reaches quiet-Sun values (Fig.~\ref{FIG13}). This
radius is inherently larger than the one based on the
intensity threshold used in identifying pores. The photometric radius roughly
corresponds to the location, where the radial velocity has a minimum (strongest
outflow) and the the divergence is zero.

There is an ongoing discussion regarding in- and outflows around pores. Many
studies \citep[e.g.,][]{Roudier2002, VargasDominguez2010} find outflows outside
and inflows inside pores. In our study, we present the exact location, where the
in- and outflows (on average) start and terminate. However, to contribute to the
debate about the existence of moat and Evershed flow around pores
\citep[e.g.,][]{VargasDominguez2008}, we need more information. Hence, the next
step is to carry out a similar study for Doppler velocities and magnetic fields.
Current simulations of magnetic features such as pores usually describe
individual pores \citep[e.g.,][]{Leka2001, Cameron2007, Rempel2011b}. It will be
very instructive, once future modeling efforts produce ensembles of pores for
different input parameters and in different magnetic settings. In principle,
these modeled pores should reproduce the mean values together with the shape of
the specific distributions for all physical parameters, presented in this
statistical study.

The results of this ongoing research are only a first step
towards a better understanding of pores and provide important information for
numerical models \citep{Leka2001, Cameron2007, Rempel2011c}. Moving forward,
SDO/HMI data offer another opportunity to study the solar cycle dependence and
temporal evolution of pores. Even though the spatial resolution is much reduced
as compared to Hinode/SOT, quasi-simultaneous intensity maps, magnetograms, and
dopplergrams cover a much broader parameter space, thus providing even tighter
boundary conditions for theoretical models. Images taken
in the extreme ultraviolet are an addtional resource provided by SDO, which
potentially links the photospheric properties of pores to the dynamics of the
Sun's upper atmosphere.

%###############################################################################
%#
%#    ACKNOWLEDGEMENTS
%#
%###############################################################################

\begin{acknowledgements}

\noindent Hinode is a Japanese mission developed and launched by ISAS/JAXA,
collaborating with NAOJ as a domestic partner, NASA and STFC (UK) as
international partners. Scientific operation of the \textit{Hinode} mission is
conducted by the Hinode science team organized at ISAS/JAXA. This team mainly
consists of scientists from institutes in the partner countries. Support for the
post-launch operation is provided by JAXA and NAOJ (Japan), STFC (UK), NASA,
ESA, and NSC (Norway). MV expresses her gratitude for the generous financial
support by the German Academic Exchange Service (DAAD) in the form of a PhD
scholarship. CD was supported by grant DE 787/3-1 of the German Science
Foundation (DFG). We would like to thank Drs. Nazaret Bello Gonz\'alez, Karin
Muglach, Horst Balthasar, and Christoph Kuckein for carefully reading the
manuscript and providing ideas,which significantly enhanced the paper.

\end{acknowledgements}

%###############################################################################
%#
%#    BIBLIOGRAPHY
%#
%###############################################################################


\begin{thebibliography}{56}
\expandafter\ifx\csname natexlab\endcsname\relax\def\natexlab#1{#1}\fi

\bibitem[{{Balthasar} {et~al.}(2000){Balthasar}, {Collados}, \&
  {Muglach}}]{Balthasar2000}
{Balthasar}, H., {Collados}, M., \& {Muglach}, K. 2000, AN, 321, 121

\bibitem[{{Balthasar} \& {G{\"o}m{\"o}ry}(2008)}]{Balthasar2008}
{Balthasar}, H. \& {G{\"o}m{\"o}ry}, P. 2008, A\&A, 488, 1085

\bibitem[{{Bray} \& {Loughhead}(1964)}]{Bray1964}
{Bray}, R.~J. \& {Loughhead}, R.~E. 1964, {Sunspots} (London: The International
  Astrophysics Series, Chapman \& Hall)

\bibitem[{{Cabrera Solana} {et~al.}(2006){Cabrera Solana}, {Bellot Rubio},
  {Beck}, \& {del Toro Iniesta}}]{CabreraSolana2006}
{Cabrera Solana}, D., {Bellot Rubio}, L.~R., {Beck}, C., \& {del Toro Iniesta},
  J.~C. 2006, ApJL, 649, L41

\bibitem[{{Cameron} {et~al.}(2007){Cameron}, {Sch{\"u}ssler}, {V{\"o}gler}, \&
  {Zakharov}}]{Cameron2007}
{Cameron}, R., {Sch{\"u}ssler}, M., {V{\"o}gler}, A., \& {Zakharov}, V. 2007,
  A\&A, 474, 261

\bibitem[{{Deng} {et~al.}(2007){Deng}, {Choudhary}, {Tritschler}, {Denker},
  {Liu}, \& {Wang}}]{Deng2007}
{Deng}, N., {Choudhary}, D.~P., {Tritschler}, A., {et~al.} 2007, ApJ, 671, 1013

\bibitem[{{Denker}(1998)}]{Denker1998b}
{Denker}, C. 1998, SoPh, 180, 81

\bibitem[{{Fanning}(2011)}]{Fanning2011}
{Fanning}, D.~W. 2011, {Coyote's Guide to Traditional IDL Graphics} (Fort
  Collins, Colorado: Coyote Book Publishing)

\bibitem[{{Gonzalez} \& {Woods}(2002)}]{Gonzalez2002}
{Gonzalez}, R.~C. \& {Woods}, R.~E. 2002, {Digital Image Processing} (Upper
  Saddle River, New Jersey: Prentice-Hall)

\bibitem[{{Hartlep} {et~al.}(2012){Hartlep}, {Busse}, {Hurlburt}, \&
  {Kosovichev}}]{Hartlep2012}
{Hartlep}, T., {Busse}, F.~H., {Hurlburt}, N.~E., \& {Kosovichev}, A.~G. 2012,
  419, 2325

\bibitem[{{Henney} {et~al.}(2009){Henney}, {Keller}, {Harvey}, {Georgoulis},
  {Hadder}, {Norton}, {Raouafi}, \& {Toussaint}}]{Henney2009}
{Henney}, C.~J., {Keller}, C.~U., {Harvey}, J.~W., {et~al.} 2009, in ASP Conf.
  Ser., Vol. 405, Solar Polarization V, ed. S.~V. {Berdyugina}, K.~N.
  {Nagendra}, \& R.~{Ramelli}, 47--50

\bibitem[{{Hirzberger}(2003)}]{Hirzberger2003}
{Hirzberger}, J. 2003, A\&A, 405, 331

\bibitem[{{Keil} {et~al.}(1999){Keil}, {Balasubramaniam}, {Smaldone}, \&
  {Reger}}]{Keil1999}
{Keil}, S.~L., {Balasubramaniam}, K.~S., {Smaldone}, L.~A., \& {Reger}, B.
  1999, ApJ, 510, 422

\bibitem[{{Keller} {et~al.}(2003){Keller}, {Harvey}, \&
  {Giampapa}}]{Keller2003}
{Keller}, C.~U., {Harvey}, J.~W., \& {Giampapa}, M.~S. 2003, in Proc. SPIE,
  Vol. 4853, Innovative Telescopes and Instrumentation for Solar Astrophysics,
  ed. S.~L. {Keil} \& S.~V. {Avakyan}, 194--204

\bibitem[{{Keppens} \& {Martinez Pillet}(1996)}]{Keppens1996}
{Keppens}, R. \& {Martinez Pillet}, V. 1996, A\&A, 316, 229

\bibitem[{{Kosugi} {et~al.}(2007){Kosugi}, {Matsuzaki}, {Sakao}, {Shimizu},
  {Sone}, {Tachikawa}, {Hashimoto}, {Minesugi}, {Ohnishi}, {Yamada}, {Tsuneta},
  {Hara}, {Ichimoto}, {Suematsu}, {Shimojo}, {Watanabe}, {Shimada}, {Davis},
  {Hill}, {Owens}, {Title}, {Culhane}, {Harra}, {Doschek}, \&
  {Golub}}]{Kosugi2007}
{Kosugi}, T., {Matsuzaki}, K., {Sakao}, T., {et~al.} 2007, SoPh, 243, 3

\bibitem[{{Langhans} {et~al.}(2005){Langhans}, {Scharmer}, {Kiselman},
  {L{\"o}fdahl}, \& {Berger}}]{Langhans2005}
{Langhans}, K., {Scharmer}, G.~B., {Kiselman}, D., {L{\"o}fdahl}, M.~G., \&
  {Berger}, T.~E. 2005, A\&A, 436, 1087

\bibitem[{{Leighton}(1969)}]{Leighton1969}
{Leighton}, R.~B. 1969, ApJ, 156, 1

\bibitem[{{Leka} \& {Skumanich}(1998)}]{Leka1998}
{Leka}, K.~D. \& {Skumanich}, A. 1998, ApJ, 507, 454

\bibitem[{{Leka} \& {Steiner}(2001)}]{Leka2001}
{Leka}, K.~D. \& {Steiner}, O. 2001, ApJ, 552, 354

\bibitem[{{Markwardt}(2009)}]{Markwardt2009}
{Markwardt}, C.~B. 2009, in ASP Conf. Ser., Vol. 411, Astronomical Data
  Analysis Software and Systems XVIII, ed. D.~A. {Bohlender}, D.~{Durand}, \&
  P.~{Dowler}, 251--254

\bibitem[{{Mathew} {et~al.}(2009){Mathew}, {Zakharov}, \&
  {Solanki}}]{Mathew2009}
{Mathew}, S.~K., {Zakharov}, V., \& {Solanki}, S.~K. 2009, A\&A, 501, L19

\bibitem[{{Montesinos} \& {Thomas}(1997)}]{Montesinos1997}
{Montesinos}, B. \& {Thomas}, J.~H. 1997, Nature, 390, 485

\bibitem[{{November} \& {Simon}(1988)}]{November1988}
{November}, L.~J. \& {Simon}, G.~W. 1988, ApJ, 333, 427

\bibitem[{{Perona} \& {Malik}(1990)}]{Perona1990}
{Perona}, P. \& {Malik}, J. 1990, IEEE Trans. Pattern Anal. Mach. Intell., 12,
  629

\bibitem[{{Pesnell} {et~al.}(2012){Pesnell}, {Thompson}, \&
  {Chamberlin}}]{Pesnell2012}
{Pesnell}, W.~D., {Thompson}, B.~J., \& {Chamberlin}, P.~C. 2012, SoPh, 275, 3

\bibitem[{{Petrie}(2012)}]{Petrie2012}
{Petrie}, G. J.~D. 2012, SoPh, 281, 577

\bibitem[{{Rempel}(2011)}]{Rempel2011c}
{Rempel}, M. 2011, ApJ, 740, 15

\bibitem[{{Rempel} \& {Schlichenmaier}(2011)}]{Rempel2011b}
{Rempel}, M. \& {Schlichenmaier}, R. 2011, Liv. Rev. Space Phy., 8, 3

\bibitem[{{Roudier} {et~al.}(2002){Roudier}, {Bonet}, \&
  {Sobotka}}]{Roudier2002}
{Roudier}, T., {Bonet}, J.~A., \& {Sobotka}, M. 2002, A\&A, 395, 249

\bibitem[{{Rucklidge} {et~al.}(1995){Rucklidge}, {Schmidt}, \&
  {Weiss}}]{Rucklidge1995}
{Rucklidge}, A.~M., {Schmidt}, H.~U., \& {Weiss}, N.~O. 1995, 273, 491

\bibitem[{{Russ}(2011)}]{Russ2011}
{Russ}, J.~C. 2011, {The Image Processing Handbook} (Boca Raton, Florida: CRC
  Press)

\bibitem[{{Sainz Dalda} {et~al.}(2012){Sainz Dalda}, {Vargas Dom{\'{\i}}nguez},
  \& {Tarbell}}]{Dalda2012}
{Sainz Dalda}, A., {Vargas Dom{\'{\i}}nguez}, S., \& {Tarbell}, T.~D. 2012,
  ApJL, 746, L13

\bibitem[{{Sankarasubramanian} \& {Rimmele}(2003)}]{Sankarasubramanian2003}
{Sankarasubramanian}, K. \& {Rimmele}, T. 2003, ApJ, 598, 689

\bibitem[{{Scherrer} {et~al.}(2012){Scherrer}, {Schou}, {Bush}, {Kosovichev},
  {Bogart}, {Hoeksema}, {Liu}, {Duvall}, {Zhao}, {Title}, {Schrijver},
  {Tarbell}, \& {Tomczyk}}]{Scherrer2012}
{Scherrer}, P.~H., {Schou}, J., {Bush}, R.~I., {et~al.} 2012, SoPh, 275, 207

\bibitem[{{Sch{\"u}ssler} {et~al.}(2003){Sch{\"u}ssler}, {Shelyag},
  {Berdyugina}, {V{\"o}gler}, \& {Solanki}}]{Schuessler2003}
{Sch{\"u}ssler}, M., {Shelyag}, S., {Berdyugina}, S., {V{\"o}gler}, A., \&
  {Solanki}, S.~K. 2003, ApJL, 597, L173

\bibitem[{{Sobotka} {et~al.}(1999){Sobotka}, {Brandt}, \&
  {Simon}}]{Sobotka1999}
{Sobotka}, M., {Brandt}, P.~N., \& {Simon}, G.~W. 1999, A\&A, 348, 621

\bibitem[{{Sobotka} \& {Roudier}(2007)}]{Sobotka2007}
{Sobotka}, M. \& {Roudier}, T. 2007, A\&A, 472, 277

\bibitem[{{Steiner} {et~al.}(2001){Steiner}, {Hauschildt}, \&
  {Bruls}}]{Steiner2001}
{Steiner}, O., {Hauschildt}, P.~H., \& {Bruls}, J. 2001, A\&A, 372, L13

\bibitem[{{Suetterlin}(1998)}]{Suetterlin1998b}
{Suetterlin}, P. 1998, A\&A, 333, 305

\bibitem[{{S{\"u}tterlin} {et~al.}(1996){S{\"u}tterlin}, {Schr{\"o}ter}, \&
  {Muglach}}]{Suetterlin1996}
{S{\"u}tterlin}, P., {Schr{\"o}ter}, E.~H., \& {Muglach}, K. 1996, SoPh, 164,
  311

\bibitem[{{S{\"u}tterlin} \& {Wiehr}(1998)}]{Suetterlin1998a}
{S{\"u}tterlin}, P. \& {Wiehr}, E. 1998, A\&A, 336, 367

\bibitem[{{Thomas} \& {Montesinos}(1993)}]{Thomas1993}
{Thomas}, J.~H. \& {Montesinos}, B. 1993, ApJ, 407, 398

\bibitem[{{Tildesley} \& {Weiss}(2004)}]{Tildesley2004}
{Tildesley}, M.~J. \& {Weiss}, N.~O. 2004, 350, 657

\bibitem[{{Tsuneta} {et~al.}(2008){Tsuneta}, {Ichimoto}, {Katsukawa}, {Nagata},
  {Otsubo}, {Shimizu}, {Suematsu}, {Nakagiri}, {Noguchi}, {Tarbell}, {Title},
  {Shine}, {Rosenberg}, {Hoffmann}, {Jurcevich}, {Kushner}, {Levay}, {Lites},
  {Elmore}, {Matsushita}, {Kawaguchi}, {Saito}, {Mikami}, {Hill}, \&
  {Owens}}]{Tsuneta2008}
{Tsuneta}, S., {Ichimoto}, K., {Katsukawa}, Y., {et~al.} 2008, SoPh, 249, 167

\bibitem[{{Uitenbroek} {et~al.}(2006){Uitenbroek}, {Balasubramaniam}, \&
  {Tritschler}}]{Uitenbroek2006}
{Uitenbroek}, H., {Balasubramaniam}, K.~S., \& {Tritschler}, A. 2006, ApJ, 645,
  776

\bibitem[{{Vargas Dom{\'{\i}}nguez} {et~al.}(2010){Vargas Dom{\'{\i}}nguez},
  {de Vicente}, {Bonet}, \& {Mart{\'{\i}}nez Pillet}}]{VargasDominguez2010}
{Vargas Dom{\'{\i}}nguez}, S., {de Vicente}, A., {Bonet}, J.~A., \&
  {Mart{\'{\i}}nez Pillet}, V. 2010, A\&A, 516, A91

\bibitem[{{Vargas Dom{\'{\i}}nguez} {et~al.}(2008){Vargas Dom{\'{\i}}nguez},
  {Rouppe van der Voort}, {Bonet}, {Mart{\'{\i}}nez Pillet}, {Van Noort}, \&
  {Katsukawa}}]{VargasDominguez2008}
{Vargas Dom{\'{\i}}nguez}, S., {Rouppe van der Voort}, L., {Bonet}, J.~A.,
  {et~al.} 2008, ApJ, 679, 900

\bibitem[{{Verma} {et~al.}(2012){Verma}, {Balthasar}, {Deng}, {Liu}, {Shimizu},
  {Wang}, \& {Denker}}]{Verma2012a}
{Verma}, M., {Balthasar}, H., {Deng}, N., {et~al.} 2012, A\&A, 538, A109

\bibitem[{{Verma} \& {Denker}(2011)}]{Verma2011}
{Verma}, M. \& {Denker}, C. 2011, A\&A, 529, A153

\bibitem[{{Verma} \& {Denker}(2012)}]{Verma2012b}
{Verma}, M. \& {Denker}, C. 2012, A\&A, 545, A92

\bibitem[{{Verma} {et~al.}(2013){Verma}, {Steffen}, \& {Denker}}]{Verma2013}
{Verma}, M., {Steffen}, M., \& {Denker}, C. 2013, A\&A, 555, A136

\bibitem[{{Villarino}(2006)}]{Villarino2006}
{Villarino}, M.~B. 2006, J. Ineq. Pure Appl. Math., 7, 21

\bibitem[{{Wang} \& {Zirin}(1992)}]{Wang1992}
{Wang}, H. \& {Zirin}, H. 1992, SoPh, 140, 41

\bibitem[{{Yang} {et~al.}(2003){Yang}, {Xu}, {Wang}, \& {Denker}}]{Yang2003a}
{Yang}, G., {Xu}, Y., {Wang}, H., \& {Denker}, C. 2003, ApJ, 597, 1190

\bibitem[{{Zuccarello} {et~al.}(2009){Zuccarello}, {Romano}, {Guglielmino},
  {Centrone}, {Criscuoli}, {Ermolli}, {Berrilli}, \& {Del
  Moro}}]{Zuccarello2009}
{Zuccarello}, F., {Romano}, P., {Guglielmino}, S.~L., {et~al.} 2009, A\&A, 500,
  L5

\end{thebibliography}
\end{document}